\documentclass[11pt,twoside,a4paper]{article}

\usepackage{epsfig}
\usepackage{amsfonts}
\usepackage{amssymb}
\usepackage{amstext}
\usepackage{amsmath}
\usepackage{xspace}
\usepackage{shadow,times}
\usepackage{algorithm,algorithmic}

\usepackage{fullpage}
\newcommand{\ignore}[1]{}

\newtheorem{thm}{Theorem}
\newtheorem{lem}[thm]{Lemma}

\newtheorem{cl}[thm]{Claim}
\newtheorem{claim}[thm]{Claim}
\newtheorem{defn}[thm]{Definition}

\newenvironment{pf}{\noindent{\bf Proof:  }}{\hfill\rule{2mm}{2mm}\medskip}
\def\sse{\subseteq}
\def\llog{\log\!\log}
\def\sse{\subseteq}

\def\cov{{\sf Partial}\xspace}

\def\lb{{\sf LB}\xspace}
\def\tsp{{\sf Tsp}}

\def\mualg{{\sf UncapMulti} }
\def\pre{{\sf PreProc} }
\def\prew{{\sf PreProc}}

\def\dem{{\sf dem} }
\def\dr{{Dial-a-Ride}\xspace}
\def\mdr{{\sf mDaR}\xspace}

\def\is{\ensuremath{\mathcal{I}}}

\title{Minimum Makespan Multi-vehicle Dial-a-Ride\thanks{A preliminary version appeared in the Proceedings of the $17^{th}$ Annual
European Symposium on Algorithms (ESA) 2009.}}
\author{Inge Li G\o rtz\thanks{Technical University of Denmark. Sponsored in part by a grant from the Carlsberg
Foundation.} \and Viswanath Nagarajan\thanks{IBM T.J. Watson Research Center.} \and R. Ravi\thanks{Tepper School of
Business, Carnegie Mellon University. Supported in part by NSF grant CCF-0728841.} }
\date{}
\begin{document}
\maketitle
\begin{abstract}
\dr problems consist of a set $V$ of $n$ vertices in a metric space (denoting travel time between vertices) and a set
of $m$ objects represented as source-destination pairs $\{(s_i,t_i)\}^m_{i=1}$, where each object requires to be moved
from its source to destination vertex. In the {\em multi-vehicle \dr} problem, there are $q$ vehicles each having
capacity $k$ and where each vehicle $j\in [q]$ has its own depot-vertex $r_j\in V$. A feasible schedule consists of a
capacitated route for each vehicle (where vehicle $j$ originates and ends at its depot $r_j$) that together move all
objects from their sources to destinations. The objective is to find a feasible schedule that minimizes the maximum
completion time (i.e. {\em makespan}) of vehicles, where the completion time of vehicle $j$ is the time when it returns
to its depot $r_j$ at the end of its route. We study the {\em preemptive} version of multi-vehicle \dr, where an object
may be left at intermediate vertices and transported by more than one vehicle, while being moved from source to
destination.
%Approximation algorithms for the single vehicle \dr problem ($q=1$) have been considered in~\cite{cr,ghnr}.
\vspace{.2cm}

Our main results are an $O(\log^3n)$-approximation algorithm for {\em preemptive multi-vehicle \dr}, and an improved
$O(\log t)$-approximation for its special case when there is no capacity constraint (here $t\le n$ is the number of
distinct depot-vertices).
%Our algorithm is based on an interesting recursive framework, and uses a new structural
%property of single-vehicle \dr tours.
%The improvement in the uncapacitated case comes from the use of sparse spanners
%in constructing the routes traced by objects.
There is an $\Omega(\log^{1/4-\epsilon} n)$ hardness of approximation known even for single vehicle capacitated
\dr~\cite{g}. For uncapacitated multi-vehicle \dr, we show that there are instances where natural lower bounds (used in
our algorithm) are $\widetilde\Omega(\log t)$ factor away from the optimum. \vspace{.2cm}

We also consider the special class of metrics induced by graphs excluding any fixed minor (eg. planar metrics). In this
case, we obtain improved guarantees of $O(\log^2n)$ for capacitated multi-vehicle \dr, and $O(1)$ for the uncapacitated
problem.
\end{abstract}

\section{Introduction}

The {\em multi-vehicle \dr} problem involves routing a set of $m$ objects from their sources to respective destinations
using a set of $q$ vehicles starting at $t$ distinct depot vertices in an $n$-vertex metric. Each vehicle has a {\em
capacity} $k$ which is the maximum number of objects it can carry at any time. Two versions arise based on whether or
not the vehicle can use any vertex in the metric as a preemption (a.k.a. transshipment) point - we study the
less-examined {\em preemptive version} in this paper. The objective in these problems is either the total completion
time or the makespan (maximum completion time) over the $q$ vehicles, and we study the more interesting {\em makespan}
version of the problem. Thus this paper studies the preemptive, capacitated, minimum makespan, multi-vehicle \dr (\mdr)
problem.

While the multiple qualifications may make the problem appear contrived, this is exactly the problem that models
courier or mail delivery over a day from several city depots: preemption is cheap and useful for packages, trucks are
capacitated and the makespan reflects the daily working time limit for each truck. Despite its ubiquity, this problem
has not been as well studied as other \dr versions. One reason from the empirical side is the difficulty in handling
the possibility of preemptions in a clean mathematical programming model. On the theoretical side which is the focus of
this paper, the difficulty of using preemption in a meaningful way in an approximation algorithm persists. It is
further compounded by the hardness of the makespan objective.

The requirement in preemptive \dr, that preemptions are allowed at all vertices, may seem unrealistic. In practice, a
subset $P$ of the vertex-set $V$ might represent  the vertices where preemption is permitted: the two extremes of this
general problem are non-preemptive \dr ($P=\emptyset$) and preemptive \dr ($P=V$). However preemptive \dr is more
generally applicable: specifically in situations where the preemption-points $P$ form a {\em net} of the underlying
metric (i.e. every vertex in $V$ has a nearby preemption-point). In particular, approximation algorithms for preemptive
\dr imply good approximations even in this general setting, wherein the precise guarantee depends on how well $P$
covers $V$.

\ignore{An example where preemptions are allowed at all vertices is a current project at a Danish hospital, where
robots in the basement will bring trash, dirty/clean linen, blood test etc.\ around. Here preemption is allowed almost
everywhere as mostly the robots will be using the basement.}
%, robots are capacitated and the makespan reflects the daily time limit.

We also note that though our model allows any number of preemptions and preemptions at all vertices, our algorithms do
not use this possibility to its full extent. Our algorithm for  the capacitated case preempts each object at most once
and our algorithm for the uncapacitated case only preempts objects at depot vertices (and at most $O(\log t)$ times).

The preemptive \dr problem has been considered earlier with a single vehicle, for which an $O(\log n)$
approximation~\cite{cr} and an $\Omega(\log^{1/4-\epsilon} n)$ hardness of approximation (for any constant $\epsilon
>0$)~\cite{g} are known. Note that the completion time and makespan objectives coincide for this case.

Moving to multiple vehicles, the total completion time objective admits a straightforward $O(\log n)$ approximation
along the lines of the single vehicle problem~\cite{cr}: Using the FRT tree embedding~\cite{frt}, one can reduce to
tree-metrics at the loss of an expected $O(\log n)$ factor, and there is a simple constant factor approximation for
this problem on trees. The maximum completion time or makespan objective, which we consider in this paper turns out to
be considerably harder. Due to non-linearity of the makespan objective, the above reduction to tree-metrics does not
hold. Furthermore, the makespan objective does not appear easy to solve even on trees.

Unlike in the single-vehicle case, note that an object in multi-vehicle \dr may be transported by several vehicles one
after the other. Hence it is important for the vehicle routes to be coordinated so that the objects trace valid paths
from respective sources to destinations. For example, a vehicle may have to wait at a vertex for other vehicles
carrying common objects to arrive. The multi-vehicle \dr problem captures aspects of both machine scheduling and
network design problems.

%\vspace{-2mm}
\subsection{Results, Techniques and Paper Outline}
\noindent{\bf Uncapacitated \mdr.} We first consider the special case of multi-vehicle \dr  where the vehicles have no
capacity constraints (i.e. $k\ge m$). This problem is interesting in itself, and serves as a good starting point before
we present the algorithm for the general case. The uncapacitated \mdr problem itself highlights differences from the
single vehicle case: For example, in single vehicle \dr, preemption plays no role in the absence of capacity
constraints; whereas in the multi-vehicle case, an optimal non-preemptive schedule may take $\Omega(\sqrt{q})$ longer
than the optimal preemptive schedule (see Section~\ref{sec:uncap-p-mdr}). Our first main result is the following:
%We prove the following theorem in Section~\ref{sec:uncap-p-mdr}.
%\vspace{-1mm}
\begin{thm}\label{th:uncap-mdr}
There is an $O(\log t)$-approximation algorithm for uncapacitated preemptive \mdr. Additionally, the schedule preempts
objects only at depot vertices.
\end{thm}
%\vspace{-1mm}
The above algorithm has two main steps: the first one (in Subsection~\ref{subsec:gen-to-dep}) reduces the instance at a
constant factor loss in the performance guarantee to one in which all demands are between depots (a ``depot-demand"
instance). In the second step (Subsection~\ref{subsec:uncap-xdepot}), we use a {\em sparse spanner} on the demand graph
to construct routes for moving objects across depots. %%

We also construct instances of uncapacitated \mdr where the optimal value is $\Omega(\log t/\log\log t)$ times all our
lower bounds for this problem (Subsection~\ref{subsec:uncap-lb}). This suggests that stronger lower bounds are needed
to obtain a better approximation ratio than what our approach provides.

We then consider the special class of metrics induced by graphs excluding some fixed minor (such as planar or
bounded-genus graphs), and obtain the following improved guarantee in Subsection~\ref{subsec:uncap-planar}.
%\vspace{-1mm}
\begin{thm}\label{th:uncap-planar}
There is an $O(1)$-approximation algorithm for uncapacitated preemptive \mdr on metrics induced by graphs that exclude
any fixed minor.
\end{thm}
Again the resulting schedule only preempts objects at depot-vertices. Furthermore, each object is preempted at most
thrice; whereas the algorithm in Theorem~\ref{th:uncap-mdr} might preempt each object $O(\log t)$ times. The algorithm
in Theorem~\ref{th:uncap-planar} has the same high-level approach outlined for Theorem~\ref{th:uncap-mdr}: the
difference is in the second step, where we use a stronger notion of {\em sparse covers} in such metrics (which follows
from the KPR decomposition theorem~\cite{kpr}), to construct routes for moving objects across depots.

%\vspace{-1mm}
\paragraph{Capacitated \mdr.} In Section~\ref{sec:p-mdr}, we study the capacitated multi-vehicle \dr problem, and obtain our second main result.
Recall that there is an $\Omega(\log^{1/4-\epsilon}n)$ hardness of approximation for even single vehicle \dr~\cite{g}.
%A feasible solution to preemptive \mdr is said to be {\em 1-preemptive} if every object is preempted at most once while being moved from its source to destination.
%\vspace{-1mm}
\begin{thm}\label{th:p-mdr}
There is a randomized $O(\log^3 n)$-approximation algorithm for preemptive \mdr. Additionally, the schedule preempts
each object at most once.
\end{thm}
%\vspace{-1mm}
This algorithm is considerably more complex than the one for the uncapacitated special case, and is the main technical
contribution of this paper. It has four key steps: (1) We {\em preprocess} the input so that demand points that are
sufficiently far away from each other can be essentially decomposed into separate instances for the algorithm to handle
independently. (2) We then solve a single-vehicle instance of the problem that obeys some additional {\em
bounded-delay} property (Theorem~\ref{th:1pmt-bnd-delay}) that we prove; This property combines ideas from algorithms
for {\em light approximate shortest path trees}~\cite{kry} and {\em capacitated vehicle routing}~\cite{hk}. The
bounded-delay property is useful in {\em randomly partitioning} the single vehicle solution among the $q$ vehicles
available to share this load. This random partitioning scheme is reminiscent of the work of Hochbaum-Maass~\cite{hm},
Baker~\cite{b} and Klein-Plotkin-Rao~\cite{kpr}, in trying to average out the effect of the cutting in the objective
function. (3) The partitioned segments of the single vehicle tour are assigned to the available vehicles; However, to
check if this assignment is feasible we solve a matching problem that identifies cases when this load assignment must
be {\em rebalanced}. This is perhaps the most interesting step in the algorithm since it identifies stronger lower
bounds for subproblems where the current load assignment is not balanced. (4) We finish up by {\em recursing} on the
load rebalanced subproblem; An interesting feature of the recursion is that the fraction of demands that are processed
recursively is not a fixed value (as is more common in such recursive algorithms) but is a carefully chosen function of
the number of vehicles on which these demands have to be served.

We prove the new bounded-delay property of single-vehicle \dr in Subsection~\ref{subsec:1-pmt}. Then we present the
algorithm for Theorem~\ref{th:p-mdr}  in Subsection~\ref{subsec:mdar-algo} and establish an $O(\log^2m\,\log
n)$-approximation bound. Using some additional preprocessing, we show in Subsection~\ref{subsec:wt-mdar} how to remove
the dependence on $m$ (number of objects) to obtain the final $O(\log^3n)$ approximation ratio.

\ignore{When the underlying metric is induced by planar graphs, we show that the structure of the sparse covers from
Busch et al.~\cite{blt} can be used to establish a stronger bounded-delay property in step~(2) of the above framework.
This immediately leads to the following improved guarantee, proved in Subsection~\ref{subsec:cap-planar}.}

When the underlying metric is induced by graphs excluding a fixed minor, we can  establish a stronger bounded-delay
property in step~(2) of the above framework. The main idea here is the construction of {\em well-separated covers} in
such metrics, which we show can be obtained using the KPR decomposition~\cite{kpr}.
This leads to the following improved guarantee, proved in Subsection~\ref{subsec:cap-planar}.%\vspace{-1mm}
\begin{thm}\label{th:cap-planar}
There is an $O(\log^2n)$-approximation algorithm for preemptive \mdr on metrics induced by graphs excluding any fixed
minor.
\end{thm}

%Due to lack of space some proofs are omitted from this paper. The proofs can be found in the full version.
%We note that in the tour created by the above algorithm each object is preempted at most once.

\ignore{In Appendix~\ref{app:eucl-mdr} we show that for Euclidean instances, better bounds can be obtained in the
structural property of Theorem~\ref{th:1pmt-bnd-delay}. This immediately implies an improved guarantee for preemptive
\mdr.
%\vspace{-1mm}
\begin{thm}\label{th:mdr-eucl}
There is an $O(\log^2 n)$-approximation algorithm for Euclidean preemptive \mdr.
\end{thm}
%\vspace{-1mm}
Finally we also consider the {\em non-preemptive} version of \mdr in Appendix~\ref{app:np-mdr}. The algorithmic
framework for the preemptive \mdr problem gives the following, which matches (up to a logarithmic factor) the best
known approximation~\cite{cr,ghnr} for the single vehicle problem.
%\vspace{-1mm}
\begin{thm}\label{th:np-mdr}
There is an $O(\sqrt{\min\{n,k\}}\cdot \log^2m)$-approximation algorithm for non-preemptive \mdr.
\end{thm}
}

%\vspace{-5mm}
\subsection{Related Work}
%\vspace{-1mm}
\dr problems form an interesting subclass of Vehicle Routing Problems that are well studied in the operations research
literature. Paepe et al.~\cite{plss} provide a classification of \dr problems using a notation similar to that for
scheduling and queuing problems: preemption is one aspect in this classification. Savelsberg and Sol~\cite{savs} and
Cordeau and Laporte~\cite{cl}
survey several variants of non-preemptive \dr problems that have been studied in the literature. %!!! We note that most
Most  \dr problems arising in practice involve making routing decisions for multiple vehicles.

\dr problems with transshipment (the preemptive version) have been studied in~\cite{ml,mp,nn}. These papers consider a
more general model where preemption is allowed only at a specified subset of vertices. Our model (and that
of~\cite{cr}) is the special case when every vertex can serve as a preemption point. It is clear that preemption only
reduces the cost of serving demands: Nakao and Nagamochi~\cite{nn} studied the maximum decrease in the optimal cost
upon introducing one preemption point. \cite{ml,mp} also model time-windows on the demands, and study heuristics and a
column-generation based approach; they also describe applications (eg. courier service) that allow for preemptions.

For single vehicle \dr, the best known approximation guarantee for the preemptive version is $O(\log n)$ (Charikar and
Raghavachari~\cite{cr}), and an $\Omega(\log^{1/4-\epsilon} n)$ hardness of approximation (for any constant $\epsilon
>0$) is shown in G\o rtz~\cite{g}. The non-preemptive version appears much harder and the best known approximation
ratio is $\min\{\sqrt{k}\log n, \sqrt{n}\log^2n\}$ (Charikar and Raghavachari~\cite{cr}, Gupta et al.~\cite{ghnr});
however to the best of our knowledge, APX-hardness is the best lower bound. There are known instances of single vehicle
\dr where the ratio between optimal non-preemptive and preemptive tours is $\Omega(\sqrt{n})$ in general
metrics~\cite{cr}, and $\tilde\Omega(n^{1/8})$ in the Euclidean plane~\cite{ghnr}. A $1.8$-approximation is known for
the $k=1$ special case of single vehicle \dr (a.k.a. {\em stacker-crane} problem)~\cite{fhk}.

Charikar et al.~\cite{ckr} studied the related {\em $k$-delivery TSP} problem, which involves transporting a number of
{\em identical objects} from supply to demand vertices, using a single capacity $k$ vehicle. The key difference from
\dr is that an object can be moved from any supply vertex to any demand vertex. \cite{ckr} gave an approximation
algorithm for $k$-delivery TSP that outputs a non-preemptive tour of length at most five times an optimal preemptive
tour. They also showed that for any $k$-delivery TSP instance, the optimal non-preemptive tour is at most four times
the optimal preemptive tour.

The {\em truck and trailer routing problem}~\cite{ch,sc} is another problem where preemption plays a crucial role. Here
a number of capacitated trucks and trailers are used to deliver all objects. Some customers are only accessible without
the trailer. The trailers can be parked at any point accessible with a trailer and it is possible to shift demand loads
between the truck and the trailer at the parking places. The papers~\cite{ch,sc} present heuristics for this problem.

%Connection to BaB..
Single vehicle preemptive \dr is closely related to the uniform {\em buy-at-bulk} problem (with cost function $\lceil
\frac{x}k\rceil$ where $k$ is the vehicle capacity). Such a connection was formally used in G{\o}rtz~\cite{g} to
establish the hardness of approximation for the single vehicle problem. Approximation algorithms for several
buy-at-bulk variants have been studied recently, eg. non-uniform buy-at-bulk~\cite{ck05,chks06}, buy-at-bulk with
node-costs~\cite{chks07} and buy-at-bulk with protection~\cite{acsz07}; poly-logarithmic approximation guarantees are
known for all these problems. However the techniques required to solve the {\em multi-vehicle} \dr problem appear quite
different from these buy-at-bulk results.

The uncapacitated \mdr problem generalizes  the {\em nurse-station-location} problem that was studied in Even et
al.~\cite{egkrs} (where a 4-approximation algorithm was given). In fact we also use this algorithm as a subroutine for
uncapacitated \mdr. Nurse-station-location is the special case of uncapacitated \mdr when each source-destination pair
coincides on a single vertex. In this paper, we handle not only the case with arbitrary pairs (uncapacitated \mdr), but
also the more general problem (capacitated \mdr) with finite capacity restriction.

\ignore{There is a large body of work on approximation algorithms for vehicle routing problems, we list some of these.
Charikar et al.~\cite{ckr} consider the problem of transporting a number of
identical objects from supply %vertices
to demand vertices. The {\em minimum latency problem}~\cite{bccprs,cgrt,jr} involves visiting locations from a depot so
as to minimize the sum of arrival times at locations. \cite{bcklmm,bbcm,nr2,ckp} study the {\em orienteering} problem:
given vertices in a metric space with associated profits, compute a bounded-length path obtaining maximum profit.
Variants of the {\em distance constrained VRP} are studied in~\cite{nr1,kmm}.}

%\vspace{-5mm}
\subsection{Problem Definition and Preliminaries}\label{subsec:prelim} We represent a finite metric as $(V,d)$ where
$V$ is the set of vertices and $d$ is a symmetric distance function satisfying the triangle inequality. For subsets
$A,B\sse V$ we denote by $d(A,B)$ the minimum distance between a vertex in $A$ and another in $B$, so $d(A,B)=\min\{
d(u,v) \mid u\in A, v\in B\}$. For a subset $E\sse {V\choose 2}$ of edges, $d(E):= \sum_{e\in E} d_e$ denotes the total
length of edges in $E$.

The {\em multi-vehicle \dr problem} (\mdr) consists of an $n$-vertex metric $(V,d)$, $m$ objects specified as
source-destination pairs $\{s_i,t_i\}_{i=1}^m$, $q$ vehicles having respective depot-vertices $\{r_j\}_{j=1}^q$, and a
common vehicle capacity $k$. A feasible schedule is a set of $q$ routes, one for each vehicle (where the route for
vehicle $j\in[q]$ starts and ends at $r_j$), such that no vehicle carries more than $k$ objects at any time and each
object is moved from its source to destination. The completion time $C_j$ of any vehicle $j\in [q]$ is the time when
vehicle $j$ returns to its depot $r_j$ at the end of its route (the schedule is assumed to start at time $0$). The
objective in \mdr is to minimize the makespan, i.e., $\min \max_{j\in [q]} C_j$. We denote by $S := \{s_i\mid i\in
[m]\}$ the set of sources, $T :=\{t_i\mid i\in [m]\}$ the set of destinations, $R :=\{r_j\mid j\in [q]\}$ the set of
distinct depot-vertices, and $t:=|R|$ the number of distinct depots. In this paper, we only consider the {\em
preemptive} version, where objects may be left at intermediate vertices while being moved from source to destination.

\paragraph{Single vehicle \dr.} The following are lower bounds for the single vehicle problem: the minimum length TSP tour on
the depot and all source/destination vertices ({\em Steiner} lower bound), and $\frac{\sum_{i=1}^m d(s_i,t_i)}{k}$
({\em flow} lower bound). Charikar and Raghavachari~\cite{cr} gave an $O(\log n)$-approximation algorithm for this
problem based on the above lower bounds. A feasible solution to preemptive \dr is said to be {\em 1-preemptive} if
every object is preempted at most once while being moved from its source to destination. Gupta et al.~\cite{ghnr}
showed that the single vehicle preemptive \dr problem always has a 1-preemptive tour of length $O(\log^2n)$ times the
Steiner and flow lower-bounds.

\paragraph{Lower bounds for \mdr.} The quantity $\frac{\sum_{i=1}^m d(s_i,t_i)}{qk}$ is a lower bound similar to
the flow bound for single vehicle \dr. Analogous to the Steiner lower bound above, is the optimal value of an induced
{\em nurse-station-location} instance. In the nurse-station-location problem~\cite{egkrs}, we are given a metric
$(V,d)$, a set $\mathcal{T}$ of terminals and a multi-set $\{r_j\}_{j=1}^q$ of depot-vertices; the goal is to find a
collection $\{F_j\}_{j=1}^q$ of trees that collectively contain all terminals $\mathcal{T}$ such that each tree $F_j$
is rooted at vertex $r_j$ and $\max_{j=1}^q d(F_j)$ is minimized. Even et al.~\cite{egkrs} gave a 4-approximation
algorithm for this problem. The optimal value of the nurse-station-location instance with depots $\{r_j\}_{j=1}^q$
(depots of vehicles in \mdr) and terminals $\mathcal{T}=S\cup T$ is a lower bound for \mdr. The following are some
lower bounds implied by nurse-station-location: (a) $1/q$ times the minimum length forest that connects every vertex in
$S\cup T$ to some depot vertex, (b) $\max_{i\in [m]}d(R,s_i)$, and (c) $\max_{i\in [m]}d(R,t_i)$. Finally, it is easy
to see that $\max_{i\in [m]} d(s_i,t_i)$ is also a lower bound for \mdr.

We note that our approximation bounds for uncapacitated \mdr are relative to the above lower bounds. However the
algorithm for capacitated \mdr relies additionally on stronger lower bounds derived from suitable subproblems.

%%%%%%%%%%%%%%%%%%%%%%% Uncapacitated %%%%%%%%%%%%%%%%%

\section{Uncapacitated Multi-Vehicle \dr}\label{sec:uncap-p-mdr}
In this section we study the uncapacitated special case of \mdr, where vehicles have no capacity constraints (i.e.
capacity $k\ge m$). We give an algorithm that achieves an $O(\log t)$ approximation ratio for this problem (recall
$t\le n$ is the number of distinct depots). Unlike in the single vehicle case, preemptive and non-preemptive versions
of \mdr are very different
even without capacity constraints.\\

\noindent {\bf Preemption gap in Uncapacitated \mdr.} Consider an instance of uncapacitated \mdr where the metric is
induced by an unweighted star with $q$ leaves (where $q$ is number of vehicles), all $q$ vehicles have the center
vertex as depot, and there is a demand between every pair of leaf-vertices. A preemptive schedule having makespan $4$
is as follows: each vehicle $j\in [q]$ visits leaf $j$ and brings all demands with source $j$ to the root, then each
vehicle $j$ visits its corresponding leaf again, this time delivering all demands with destination $j$. On the other
hand, in any non-preemptive schedule, one of the $q$ vehicles completely serves at least $q-1$ demands (since there are
$q(q-1)$ demands in all). The minimum length of any tour containing the end points of $q$ demands is
$\Omega(\sqrt{q})$, which is also a lower bound on the optimal non-preemptive makespan. Thus there is an
$\Omega(\sqrt{q})$ factor gap between optimal preemptive and non-preemptive tours. This is in contrast to the {\em
uncapacitated single vehicle} case, where it is easy to see that the optimal preemptive and non-preemptive tours
coincide.

\medskip

The algorithm for uncapacitated \mdr proceeds in two stages. Given any instance, it is first reduced (at the loss of a
constant factor) to a depot-demand instance, where all demands are between depot vertices
(Subsection~\ref{subsec:gen-to-dep}). This reduction uses the nurse-station-location algorithm from Even et
al.~\cite{egkrs}. Then the depot-demand instance is solved using an $O(\log t)$-approximation algorithm
(Subsection~\ref{subsec:uncap-xdepot}); this is the key step in the algorithm, and is based on a sparse-spanner on the
demand graph.

\subsection{Reduction to Depot-demand Instances}\label{subsec:gen-to-dep} We define {\em depot-demand instances} as those instances of uncapacitated \mdr where all demands are
between depot vertices. Given any instance $\mathcal{I}$ of uncapacitated \mdr, the algorithm \mualg
(Figure~\ref{fig:algo-uncap}) reduces $\mathcal{I}$ to a depot-demand instance. We now argue that the reduction in
\mualg only loses a constant approximation factor. Let $B$ denote the optimal makespan of instance $\mathcal{I}$. Since
the optimal value of the nurse-station-location instance solved in the first step of \mualg is a lower bound for
$\mathcal{I}$, we have $\max_{j=1}^q d(F_j)\le 4B$.

\begin{figure}[h]
\begin{small}\begin{center} \framebox[5.8in]{\parbox{5.5in}{ {\bf Input: instance $\mathcal{I}$ of uncapacitated \mdr.}
\begin{itemize}
\item Solve the nurse-station-location instance with depots $\{r_j\}_{j=1}^q$ and all sources/destinations $S\cup T$ as terminals, using the 4-approximation algorithm~\cite{egkrs}.
Let $\{F_j\}_{j=1}^q$ be the resulting trees covering $S\cup T$ such that each tree $F_j$ is rooted at depot $r_j$.
\item Define a depot-demand instance $\mathcal{J}$ of uncapacitated \mdr on the same metric and set of vehicles, where \ignore{the metric is induced on only the depot
vertices $\{r_j\}_{j=1}^q$,} the demands are $\{(r_j,r_l)\mid s_i\in F_j~\&~t_i\in F_l,~1\le i\le m\}$.\ignore{, and
there is one vehicle located at each $\{r_j\}_{j=1}^q$} For any object $i\in [m]$ let the \emph{source depot} be the
depot $r_j$ for which $s_i \in F_j$ and the \emph{destination depot} be the depot  $r_l$ for which $t_i \in F_l$.
\item Output the following schedule for $\mathcal{I}$:
\begin{enumerate}
 \item Each vehicle $j\in[q]$ traverses tree $F_j$ by an Euler tour, picks up all objects from sources in $F_j$ and brings them to their source-depot $r_j$.
 \item Vehicles implement a schedule for {\em depot-demand instance} $\mathcal{J}$, and all objects are moved from their source-depot to destination-depot (using the algorithm in Section~\ref{subsec:uncap-xdepot}).
 \item Each vehicle $j\in[q]$ traverses tree $F_j$ by an Euler tour, picks up all objects having destination-depot $r_j$ and brings them to their destinations in $F_j$.
\end{enumerate}
\end{itemize}
}} \caption{Algorithm \mualg for uncapacitated \mdr.} \label{fig:algo-uncap}
\end{center}\end{small}
\end{figure}

\begin{cl} \label{cl:redn-depot-inst} The optimal makespan for the depot-demand instance $\mathcal{J}$ is at most $17B$.
\end{cl}
\begin{pf}
Consider a feasible schedule for $\mathcal{J}$ involving three rounds: (1) each vehicle traverses (by means of an Euler
tour) its corresponding tree in $\{F_j\}_{j=1}^q$ and moves each object $i$ from its source-depot (the source in
instance $\mathcal{J}$) to $s_i$ (source in original instance $\mathcal{I}$); (2) each vehicle follows the optimal
schedule for $\mathcal{I}$ and moves each object $i$ from $s_i$ to $t_i$ (destination in $\mathcal{I}$); (3) each
vehicle traverses its corresponding tree in $\{F_j\}_{j=1}^q$ and moves each object $i$ from $t_i$ to its
destination-depot (the destination in $\mathcal{J}$). Clearly this is a feasible schedule for $\mathcal{J}$. From the
observation on the nurse-station-location instance, the time taken in each of the first and third rounds is at most
$8B$. Furthermore, the time taken in the second round is the optimal makespan of $\mathcal{I}$ which is $B$. This
proves the claim.
\end{pf}

Assuming a feasible schedule for $\mathcal{J}$, it is clear that the schedule returned by \mualg is feasible for the
original instance $\mathcal{I}$. The first and third rounds in $\mathcal{I}$'s schedule require at most $8B$ time each.
Thus an approximation ratio $\alpha$ for depot-demand instances implies an approximation ratio of $17\alpha + 8$ for
general instances. In the next subsection, we show an $O(\log t)$-approximation algorithm for depot-demand instances
(here $t$ is the number of depots), which implies Theorem~\ref{th:uncap-mdr}.

\subsection{Algorithm for Depot-demand Instances \label{subsec:uncap-xdepot}} Let $\mathcal{J}$ be any depot-demand instance: note that the instance defined in the
second step of \mualg is of this form. It suffices to restrict the algorithm to the induced metric $(R,d)$ on only
depot vertices, and use only one vehicle at each depot in $R$. Consider an undirected graph $H$ consisting of vertex
set $R$ and edges corresponding to demands: there is an edge between vertices $r$ and $s$ iff there is an object going
from either $r$ to $s$ {\em or} $s$ to $r$. Note that the metric length of any edge in $H$ is at most the optimal
makespan $\widetilde{B}$ of instance $\mathcal{J}$. In the schedule produced by our algorithm, vehicles will only use
edges of $H$. Thus in order to obtain an $O(\log t)$ approximation, it suffices to show that each vehicle only
traverses $O(\log t)$ edges. Based on this, we further reduce $\mathcal{J}$ to the following instance $\mathcal{H}$ of
uncapacitated \mdr: the underlying metric is shortest paths in graph $H$ (on vertices $R$), with one vehicle at each
$R$-vertex, and for every edge $(u,v)\in H$ there is a demand from $u$ to $v$ and one from $v$ to $u$. Clearly any
schedule for $\mathcal{H}$ having makespan $\beta$ implies one for $\mathcal{J}$ of makespan $\beta\cdot
\widetilde{B}$. The next lemma implies an $O(\log |R|)$ approximation for depot-demand instances.
\begin{lem} \label{lem:uncap-depot-inst} There exists a poly-time computable schedule for $\mathcal{H}$
with makespan $O(\log t)$, where $t=|R|$.
\end{lem}
\begin{pf}
Let $\alpha=\lceil \lg t\rceil +1$. We first construct a {\em sparse spanner} $A$ of $H$ as follows: consider edges of
$H$ in an arbitrary order, and add an edge $(u,v)\in H$ to $A$ iff the shortest path between $u$ and $v$ using current
edges of $A$ is more than $2\alpha$. It is clear from this construction that the girth of $A$ (length of its shortest
cycle) is at least $2\alpha$, and that for every edge $(u,v)\in H$, the shortest path between $u$ and $v$ in $A$ is at
most $2\alpha$.

We now assign each edge of $A$ to one of its end-points such that each vertex is assigned at most two edges. Repeatedly
pick any vertex $v$ of degree at most two in $A$, assign its adjacent edges to $v$, and remove these edges and $v$ from
$A$. We claim that at the end of this procedure (when no vertex has degree at most 2), all edges of $A$ would have been
removed (i.e. assigned to some vertex). Suppose for a contradiction that this is not the case. Let $\tilde{A}\ne
\emptyset$ be the remaining graph; note that $\tilde{A}\sse A$, so girth of $\tilde{A}$ is at least $2\alpha$. Every
vertex in $\tilde{A}$ has degree at least 3, and there is at least one such vertex $w$. Consider performing a
breadth-first search in $\tilde{A}$ from $w$. Since the girth of $\tilde{A}$ is at least $2\alpha$, the first $\alpha$
levels of the breadth-first search is a tree. Furthermore every vertex has degree at least 3, so each vertex in the
first $\alpha-1$ levels has at least 2 children. This implies that $\tilde{A}$ has at least $1+2^{\alpha-1}>t$
vertices, which is a contradiction! For each vertex $v\in R$, let $A_v$ denote the edges of $A$ assigned to $v$ by the
above procedure; we argued that $\cup_{v\in R} A_v = A$, and $|A_v|\le 2$ for all $v\in R$.

The schedule for $\mathcal{H}$ involves $2\alpha$ rounds as follows. In each round, every vehicle $v\in R$ traverses
the edges in $A_v$ (in both directions) and returns to $v$. Since $|A_v|\le 2$ for all vertices $v$, each round takes 4
units of time; so the makespan of this schedule is $8\alpha=O(\log t)$. The route followed by each object in this
schedule is the shortest path from its source to destination in spanner $A$; note that the length of any such path is
at most $2\alpha$. To see that this is indeed feasible, observe that every edge of $A$ is traversed by some vehicle in
each round. Hence in each round, every object traverses one edge along its shortest path (unless it is already at its
destination). Thus after $2\alpha$ rounds, all objects are at their destinations. \end{pf}

Note that the final algorithm for uncapacitated \mdr preempts each object at most $O(\log t)$ times, and only at
depot-vertices. This completes the proof of Theorem~\ref{th:uncap-mdr}.

\subsection{Tight Example for Uncapacitated \mdr Lower Bounds}\label{subsec:uncap-lb}
We note that known lower bounds for uncapacitated \mdr are insufficient to obtain a sub-logarithmic approximation
guarantee. The lower bounds we used in our algorithm are the following: $\max_{i\in [m]} d(s_i,t_i)$, and the optimal
value of a nurse-station-location instance with depots $\{r_j\}_{j=1}^q$ and terminals $S\cup T$. We are not aware of
any lower bounds stronger than these two bounds.

We show an instance $\mathcal{G}$ of uncapacitated \mdr where the optimal makespan is a factor $\Omega(\frac{\log
t}{\llog t})$ larger than both the above lower bounds. In fact, the instance we construct is a depot-demand instance
that has the same special structure as instance $\mathcal{H}$ in Lemma~\ref{lem:uncap-depot-inst}. I.e. the demand
graph is same as the graph inducing distances. Take $G=(V,E)$ to be a $t$-vertex regular graph of degree $\sim \log t$
and girth $g\sim \log t/ \llog t$ (there exist such graphs, eg. Lazebnik et al.~\cite{luw}). Instance $\mathcal{G}$ is
defined on a metric on vertices $V$ with distances being shortest paths in graph $G$. For every edge $(u,v)\in E$ of
graph $G$, there is an object with source $u$ and destination $v$ (the direction is arbitrary). There is one vehicle
located at every vertex of $V$; so number of vehicles $q=t$.

Observe that both our lower bounds are $O(1)$: the optimal value of the nurse-station-location instance is $0$, and
maximum source-destination distance is 1. However as we show below, the optimal makespan for this instance is at least
$g-1=\Omega(\log t/ \llog t)$. Suppose (for contradiction) that there is a feasible schedule for $\mathcal{G}$ with
makespan $M< g-1$. A demand $(u,v)\in E$ is said to be {\em completely served} by a vehicle $j$ iff the route of
vehicle $j$ visits both vertices $u$ and $v$. The number of distinct vertices visited by any single vehicle is at most
$M<g$: so the number of demands that are completely served by a single vehicle is at most $M-1$ (otherwise these demand
edges would induce a cycle smaller than the girth $g$). Hence the number of demands that are completely served by some
vehicle is at most $t\cdot g < |E|$. Let $(u,v)\in E$ be a demand that is {\em not} completely served by any vehicle,
i.e. there is no vehicle that visits both $u$ and $v$. Since we have a feasible schedule of makespan $M$, the path
$\pi$ followed by demand $(u,v)$ from $u$ to $v$ (or vice versa) in the schedule has length at most $M$. The path $\pi$
can not be the direct edge $(u,v)$ since demand $(u,v)$ is not completely served by any vehicle. So path $\pi$ together
with edge $(u,v)$ is a cycle of length at most $M+1<g$ in graph $G$, contradicting girth of $G$.

\subsection{Improved Algorithm for Metrics Excluding a Fixed Minor}\label{subsec:uncap-planar}
We now prove Theorem~\ref{th:uncap-planar} that gives a constant approximation algorithm for
uncapacitated \mdr on metrics induced by $K_r$-minor free graphs (for any fixed $r$). This improvement comes from using
the existence of good `sparse covers' in such metrics, as opposed to the spanner based construction in
Lemma~\ref{lem:uncap-depot-inst}. This guarantee is again relative to the above mentioned lower bounds.

Consider an instance of uncapacitated \mdr on metric $(V,d)$ that is induced by an edge-weighted graph $G=(V,E)$
containing no $K_r$-minor (for some fixed $r\ge 1$). We start with some definitions~\cite{blt}. A {\em cluster} is any
subset of vertices. For any $\gamma>0$ and vertex $v\in V$, $N(v,\gamma):=\{u\in V\mid d(u,v)\le \gamma\}$ denotes the
set of vertices within distance $\gamma$ from $v$. As observed in Busch et al.~\cite{blt} and Abraham et
al.~\cite{igmw}, the partitioning scheme of Klein et al.~\cite{kpr} implies the following result.
\begin{thm}[\cite{kpr}] \label{th:sparse-cover}Given $K_r$-minor free graph $G=(V,E,w)$ and value $\gamma>0$, there is an algorithm that computes a set $Z=\{C_1,\cdots,C_l\}$ of clusters satisfying:
\begin{enumerate}
 \item The diameter of each cluster is at most $O(r^2)\cdot\gamma$, i.e. $\max_{u,v\in C_i} d(u,v)\le O(r^2)\cdot\gamma$ for all $i\in [l]$.
 \item For every $v\in V$, there is some cluster $C_i\in Z$ such that $N(v,\gamma)\sse C_i$.
 \item For every $v\in V$, the number of clusters in $Z$ that contain $v$ is at most $O(2^r)$.
\end{enumerate}
\end{thm}
The set of clusters $Z$ found above is called a {\em sparse cover} of $G$.
\\

%\begin{thm}
%There is an $O(1)$-approximation algorithm for the uncapacitated preemptive \mdr problem on metrics induced by graphs
%that exclude any fixed minor.
%\end{thm}
\begin{pf}{\bf [Theorem~\ref{th:uncap-planar}]}
The reduction in Section~\ref{sec:uncap-p-mdr} implies that it suffices to consider depot-demand instances: An $O(1)$
approximation for such instances implies an $O(1)$ approximation for general instances. Let $\mathcal{J}$ be any
depot-demand instance on metric $(V,d)$ induced by $K_r$-minor free graph $G=(V,E)$, with a set $R\sse V$ of
depot-vertices (each containing a vehicle), and where all demands $\{s_i,t_i\}_{i=1}^m$ are between vertices of $R$.
The algorithm is described in Figure~\ref{fig:algo-plane-uncap}.
\begin{figure}[h]
\begin{center}
\framebox[6in]{\parbox{5.8in}{
\begin{small}
{\bf Input:} Depot-demand instance $\mathcal{J}$ on metric $G=(V,E,w)$, depot-vertices $R\sse V$, demands
$\{s_i,t_i\}_{i=1}^m$.
\begin{enumerate}
 \item Let $\gamma=\max_{i\in [m]} d(s_i,t_i)$ be the maximum source-destination distance.
 \item Compute a sparse cover $Z=\{C_j\}_{j=1}^l$ given in Theorem~\ref{th:sparse-cover} for parameter $\gamma$.
 \item For each cluster $C_j\in Z$, choose an arbitrary vertex $c_j\in C_j$ as its {\em center}.
 \item \label{step:planar-uncap:1}For each demand $i\in [m]$, let $\pi(i)\in [l]$ be such that $s_i,t_i\in C_{\pi(i)}$.
 \item Output the following schedule for $\mathcal{J}$:
 \begin{enumerate}
  \item \label{step:uncap-plane1} Each vehicle $r\in R$ visits the centers of all clusters containing $r$, and returns to $r$. Vehicle $r$ carries all the objects $\{i\in[m]\mid s_i=r\}$ having source $r$, and drops each object $i$ at
  center $c_{\pi(i)}$.
  \item \label{step:uncap-plane2} Each vehicle $r\in R$ again visits the centers of all clusters containing $r$. Vehicle $r$ brings all the objects $\{i\in[m]\mid t_i=r\}$ having destination $r$: each object $i$ is picked up
  from  center $c_{\pi(i)}$.
 \end{enumerate}
\end{enumerate}
{\bf Output:} An $O(1)$-approximate minimum makespan schedule for $\mathcal{J}$.
\end{small}}}
\end{center}
\caption{Algorithm for uncapacitated \mdr on $K_r$-minor free graphs.\label{fig:algo-plane-uncap}}
\end{figure}

Note that $\gamma$ is a lower bound on the optimal makespan of $\mathcal{J}$. We claim that the makespan of the above
schedule is at most $O(r^2\,2^r)\cdot \gamma$. Observe that each depot is contained in at most $O(2^r)$ clusters, and
the distance from any depot to the center of any cluster containing it is at most $O(r^2)\cdot\gamma$. Hence the time
taken by each vehicle in either of the two rounds (Steps~\eqref{step:uncap-plane1}-\eqref{step:uncap-plane2}) is at
most $O(r^2 2^r)\cdot \gamma$. Since $r$ is a fixed constant, the final makespan is $O(1)\cdot \gamma$.

We now argue the feasibility of the above schedule. Step~\eqref{step:planar-uncap:1} is well-defined: for all
$i\in[m]$, we have $s_i, t_i\in N(s_i,\gamma)$ and there is some $j\in[l]$ with $N(s_i,\gamma)\sse C_j$ (i.e. we can
set $\pi(i)=j$). It is now easy to see that each object $i\in [m]$ traces the following route in the final schedule:
$s_i\leadsto c_{\pi(i)}\leadsto t_i$.
\end{pf}

Combining this algorithm for depot-demand instances with the reduction in Subsection~\ref{subsec:gen-to-dep}, we obtain
an $O(r^2\,2^r)$-approximation algorithm for uncapacitated \mdr on $K_r$-minor free metrics. When $r$ is a fixed
constant this is a constant approximation, and we obtain Theorem~\ref{th:uncap-planar}. Note that each object in the
resulting schedule is preempted at most thrice, and only at depot-vertices.
%----------------------------Capacitated mDaR------------------------------

\section{Capacitated Multi-Vehicle \dr}\label{sec:p-mdr}
In this section we prove our main result: an $O(\log^3 n)$-approximation algorithm for the \mdr problem (with
capacities). We first obtain a new structure theorem on single-vehicle \dr tours (Subsection~\ref{subsec:1-pmt}) that
preempts each object at most once, and where the total time spent by objects in the vehicle is bounded. Obtaining such
a single vehicle tour is crucial in our algorithm for capacitated \mdr, which appears in
Section~\ref{subsec:mdar-algo}. There we prove a weaker approximation bound of $O(\log^2m\,\log n)$; in
Subsection~\ref{subsec:wt-mdar} we show how to remove the dependence on $m$ to obtain an $O(\log^3 n)$-approximation
algorithm even for ``{\em weighted} \mdr'' (i.e. Theorem~\ref{th:p-mdr}). In Subsection~\ref{subsec:cap-planar}, we
establish a stronger structure theorem (analogous to the one in Subsection~\ref{subsec:1-pmt}) for metrics excluding
any fixed minor; this immediately leads to an improved $O(\log^2n)$-approximation ratio for \mdr on such metrics (i.e.
Theorem~\ref{th:cap-planar}).

\ignore{{\bf (1)} Obtain a {\em single-vehicle} tour satisfying 1-preemptive and bounded-delay properties
(Theorem~\ref{th:1pmt-bnd-delay}), {\bf (2)} Randomly partition the single vehicle tour into $|Q|$ equally spaced
pieces, {\bf (3)} Solve a matching problem to assign {\em some} of these pieces to vehicles of $Q$ that satisfy a
subset of demands $D$, {\bf (4)} A suitable fraction of the unsatisfied demands in $D$ are covered recursively by
unused vehicles of $Q$. }

\subsection{Capacitated Vehicle Routing with Bounded Delay} \label{subsec:1-pmt}
Before we present the structural result on \dr tours, we consider the classic {\em capacitated vehicle routing
problem}~\cite{hk} with an additional constraint on object `delays'. %Recall from Chapter~\ref{chap:sdar} that the
The capacitated vehicle routing problem (CVRP) is a special case of single vehicle \dr when all objects have the same
source (or equivalently, same destination). Formally, we are given a metric $(V,d)$, specified depot-vertex $r\in V$,
and $m$ objects all having source $r$ and respective destinations $\{t_i\}_{i\in [m]}$. The goal is to compute a
minimum length {\em non-preemptive} tour of a capacity $k$ vehicle originating at $r$ that moves all objects from $r$
to their destinations. In {\em CVRP with bounded delay}, we are additionally given a {\em delay parameter} $\beta> 1$,
and the goal is to find a minimum length capacitated non-preemptive tour serving all objects such that the time spent
by each object $i\in [m]$ in the vehicle is at most $\beta\cdot d(r,t_i)$. Well-known lower bounds for the
CVRP~\cite{hk} are as follows: minimum length TSP tour on $\{r\}\cup\{t_i\mid i\in [m]\}$ (Steiner lower bound), and
$\frac{2}{k}\sum_{i=1}^m d(r,t_i)$ (flow lower bound). These lower bounds also hold for the (less constrained)
preemptive version of CVRP.
% (where objects may be left at intermediate vertices).
\begin{thm}\label{th:bnd-delay}
There is a $(2.5+\frac{3}{\beta-1})$-approximation algorithm for CVRP with bounded delay, where $\beta>1$ is the delay
parameter. This guarantee is relative to the Steiner and flow lower bounds.\ignore{Let $\mathcal{I}$ be an instance of
{\em single source} preemptive Dial-a-Ride on metric $(V,d)$ with all objects having source $r$ and respective
destinations $\{t_i\}_{i\in D}$, and let $\lb$ denote the flow/Steiner lower bound for $\mathcal{I}$. For any
$\beta>1$, there exists a poly-time computable non-preemptive tour of length at most $(2.5+\frac{3}{\beta-1})\cdot \lb$
such that the time spent by each object $i\in D$ in the vehicle is at most $\beta\cdot d(r,t_i)$.}
\end{thm}
\begin{pf} Our algorithm is basically a combination of the algorithms for {\em light approximate shortest path trees}~\cite{kry}, and capacitated vehicle
routing~\cite{hk}. Let $\lb$ denote the maximum of the Steiner and flow lower bounds. The minimum TSP tour on the
destinations plus $r$ is the Steiner lower bound. The first step is to compute an approximately minimum TSP tour $C$:
Christofides' algorithm~\cite{c} gives a 1.5-approximation, so $d(C)\le 1.5\cdot \lb$. Number the vertices in the order
in which they appear in $C$, starting with $r$ being 0. Using the procedure in Khuller et al.~\cite{kry}, we obtain a
set of edges $\{(0,v_1),\cdots ,(0,v_t)\}$ with $0<v_1< v_2< \cdots < v_t<|V|$ having the following properties (below
$v_0=0$).
\begin{enumerate}
 \item For $1\le p\le t$, for any vertex $u$ with $v_{p-1}\le u <v_p$, the length of
 edge $(0,v_{p-1})$ plus the path along $C$ from $v_{p-1}$ to $u$ is
 at most $\beta\cdot d(0,u)$.
 \item $\sum_{p=1}^t d(0,v_p) \le \frac{1}{\beta-1} \cdot d(C)$.
\end{enumerate}

For each $1\le p\le t$, define tour $C_p$ which starts at $r$, goes to $v_{p-1}$, traverses $C$ until $v_p$, then
returns to $r$. Assign vertices $\{v_{p-1},\cdots ,v_p -1\}$ (and all demands contained in them) to $C_p$. Also define
tour $C_{t+1}$ which starts at $r$, goes to $v_{t}$, and traverses $C$ until $r$; and assign all remaining demands to
$C_{t+1}$. It is clear that $C_p$ (for $1\le p\le t+1$) visits each vertex assigned to it within $\beta$ times the
shortest path from $r$ (using property~1 above). Also, the total length $\sum_{p=1}^{t+1} d(C_p) = d(C) + 2\sum_{p=1}^t
d(0,v_p) \le (1+\frac{2}{\beta-1}) d(C)$, by property~2.

For each $C_p$, we service the set $D_p$ of demands assigned to it separately. \ignore{We use the shifting argument in
the algorithm for capacitated vehicle routing~\cite{hk}.} Index the demands in $D_p$ in the order in which they appear
on $C_p$ (breaking ties arbitrarily). Consider a capacitated tour which serves these demands in groups of at most $k$
each, and returns to $r$ after serving each group. The groups are defined as follows: starting at index 1, each group
contains the next $k$ contiguous demands (until all of $D_p$ is assigned to groups). By rotating the indexing of
demands, there are $k$ different groupings of $D_p$ that can be obtained: each of which corresponds to a capacitated
tour. As argued in~\cite{hk} (and is easy to see), the average length of these $k$ tours is at most $d(C_p) +
2\sum_{z\in D_p} \frac{d(r,z)}{k}$. So the minimum length tour $\gamma_p$ among these satisfies $d(\gamma_p)\le d(C_p)
+ 2\sum_{z\in D_p} \frac{d(r,z)}{k}$.

The final solution $\gamma$ is the concatenation of tours $\gamma_1,\cdots ,\gamma_{t+1}$. Note that the time spent in
the vehicle by any demand $i$ is at most $\beta \cdot d(r,t_i)$. The length of tour $\gamma$ is at most
$\sum_{p=1}^{t+1} d(C_p) + 2\sum_{z\in D} \frac{d(r,z)}{k}$, where $D$ is the set of all demands. Note that
$2\sum_{z\in D} \frac{d(r,z)}{k}$ is the flow lower bound for this CVRP instance, so it is at most $\lb$. Hence we
obtain the following.
$$d(\gamma)\le (1+\frac{2}{\beta-1}) d(C) + \lb \le
(1+\frac{2}{\beta-1}) \frac{3}{2}\lb + \lb = (2.5+\frac{3}{\beta-1})\lb$$ Clearly, solution $\gamma$ satisfies the
desired conditions in the theorem.\end{pf}

We now consider the {\em single vehicle} preemptive \dr problem given by a metric $(V,d)$, a set $D$ of demand-pairs, and a
vehicle of capacity $k$. Given Theorem~\ref{th:bnd-delay}, the following structural result is a simple extension of
Theorem~16 from Gupta et al.~\cite{ghnr}. We present the proof for completeness.
%This is obtained immediately by using Theorem~\ref{th:bnd-delay} in the proof of Theorem~\ref{1-pmt}, in place of the ~\cite{hk} algorithm.
\begin{thm}\label{th:1pmt-bnd-delay}
There is a randomized poly-time computable 1-preemptive tour $\tau$ servicing $D$ that satisfies the following
conditions (where $\lb$ is the average of the Steiner and flow lower bounds):
\begin{enumerate}
  \item {\bf Total length:} $d(\tau)\le O(\log^2n)\cdot \lb$.
  \item {\bf Bounded delay:} $\sum_{i\in D} T_i\le O(\log n)\sum_{i\in D} d(s_i,t_i)$ where
$T_i$ is the total time spent by object $i\in D$ in the vehicle under the schedule given by $\tau$.
\end{enumerate}
\end{thm}

\begin{pf}
We follow the algorithm of Gupta et al.~\cite{ghnr} that obtains a single vehicle 1-preemptive tour within $O(\log^2
n)$ factor of the Steiner and flow lower bounds. In addition, we will also ensure the bounded-delay property.
\ignore{Lemma~\ref{lem:1-pmt} has an additional condition on the 1-preemptive tour that the total time spent by each
object in the vehicle is bounded, which is not implied by the algorithm in~\cite{ghnr}. Achieving this requires some
more work (Lemma~\ref{lem:bnd-delay}), and we give the entire proof below (the first part of the proof is included
from~\cite{ghnr} for completeness).} Using the results on probabilistic tree embedding~\cite{frt}, we may assume that
the given metric is a {\em hierarchically well-separated} tree $\mathcal{T}$ with distance function $\kappa$. This only
increases the expected value of the lower bound by a factor of $O(\log n)$; i.e. $E[\widetilde\lb] = O(\log n)\,\lb$
where $\widetilde\lb$ (resp. $\lb$) equals the average of the Steiner and flow lower bounds on metric $\kappa$ (resp.
$d$). Furthermore, as discussed in~\cite{ghnr}, we may assume that the metric $\kappa$ is induced by a tree
$\mathcal{T}$ on the original vertex set $V$ having $l=O(\log n)$ levels.
%expected length of the optimal preemptive tour (on metric $\kappa$) $\widetilde{\opt}=O(\log n)\cdot \lb$ (here $\lb$ is the preemptive lower
%bound on the original metric $d$).

We now partition the demands in $\mathcal{T}$ into $l$ sets with $D_p$ (for $p=1,\cdots,l$) consisting of all demands
$i\in D$ having their nearest common ancestor (nca) in level $p$ (i.e. the nca of $s_i$ and $t_i$ is a vertex in level
$p$). We service each $D_p$ separately using a tour of length $O(\widetilde{\lb})$. Finally we concatenate the tours
for each level $p$, to obtain the theorem.

\smallskip

\noindent {\bf Servicing $D_p$.} For each vertex $v$ at level $p$ in $\mathcal{T}$, let $L_v$ denote the demands in
$D_p$ that have $v$ as their nca. Let $\lb(v)$ denote the average of the Steiner and flow lower-bounds (in metric
$\kappa$) for the single vehicle problem with depot $v$ and demands $L_v$. Clearly the flow lower bounds are disjoint
for different vertices $v$. Also, the subtrees under any two different level $p$ vertices are disjoint, so the Steiner
lower bounds are disjoint for different $v$. Thus $\sum_v \lb(v) \le \widetilde{\lb}$. We now show how each $L_v$ is
served separately.

\smallskip

\noindent {\bf Servicing $L_v$.} Consider the following two instances of the CVRP problem in metric $\kappa$:
$\is_{src}$ with all sources in $L_v$ and common destination $v$, and $\is_{dest}$ with common source $v$ and all
destinations in $L_v$. Observe that the Steiner and flow lower bounds for each of $\is_{src}$ and $\is_{dest}$ are all
at most $2\,\lb_v$.

Consider instance $\is_{src}$. Setting delay parameter $\beta=2$ in Theorem~\ref{th:bnd-delay}, we obtain a {\em
non-preemptive} tour $\sigma_v$ that moves all objects in $L_v$ from their sources to vertex $v$, such that (1) the
length of $\sigma_v$ is at most $11\,\lb_v$, and (2) the time spent by each object $i\in L_v$ in the vehicle is at most
$2\,\kappa(s_i,v)$.  Similarly for instance $\is_{dest}$, we obtain a {\em non-preemptive} tour $\tau_v$ that moves all
objects in $L_v$ from vertex $v$ to their destinations, such that (1) the length of $\tau_v$ is at most $11\,\lb_v$,
and (2) the time spent by each object $i\in L_v$ in the vehicle is at most $2\,\kappa(v,t_i)$. Concatenating these two
tours, we obtain that $\sigma_v\cdot \tau_v$ is a {\em 1-preemptive} tour servicing $L_v$, of length at most $22\cdot
\lb(v)$, where each object $i\in L_v$ spends at most $2(\kappa(s_i,v)+\kappa(v,t_i))=2\cdot\kappa(s_i,t_i)$ time in the
vehicle.

\smallskip

\noindent We now obtain a depth-first-search traversal (DFS) on $\mathcal{T}$ (restricted to the end-points of demands
$D_p$) to visit all vertices $v$ in level $p$ that have some demand in their subtree (i.e. $L_v\ne \emptyset$), and use
the algorithm described above for servicing demands $L_v$ when $v$ is visited in the DFS. This is the tour servicing
$D_p$. Note that the length of the DFS is at most the Steiner lower bound on $\kappa$, so it is at most
$2\,\widetilde{\lb}$. Thus the tour servicing $D_p$ has length at most $2\,\widetilde{\lb} + 22\sum_v \lb(v)$, where
$v$ ranges over all vertices in level $p$ of $\mathcal{T}$. Recall that $\sum_v \lb(v) \le \widetilde{\lb}$, so the
tour servicing $D_p$ has length at most $24\cdot \widetilde{\lb}$. Additionally, the time spent by each object $i\in
D_p$ in the vehicle is at most $2\,\kappa(s_i,t_i)$.

\smallskip

\noindent Finally concatenating the tours for each level $p=1,\cdots,l$, we obtain a 1-preemptive tour on $\mathcal{T}$
of length $O(\log n)\cdot \widetilde{\lb}$. Additionally, the time spent by each object $i$ in the vehicle is at most
$2\cdot\kappa(s_i,t_i)$. This translates to a 1-preemptive tour on the original metric having {\em expected} length
$O(\log^2 n)\cdot \lb$, and with $E[\sum_{i\in D} T_i]\le \sum_{i\in D} 2\cdot E[\kappa(s_i,t_i)]\le O(\log
n)\sum_{i\in D} d(s_i,t_i)$, where $T_i$ is the time spent by object $i$ in the vehicle. Using Markov inequality and a
union bound, we can ensure that the resulting 1-preemptive tour on metric $d$ satisfies {\em both} conditions~1-2 with
constant probability.\end{pf}

\ignore{Now consider an optimal preemptive tour $\tau_v$ servicing $L_v$. Since the $s_j-t_j$ path of each demand $j\in
L_v$ crosses vertex $v$, at some point in tour $\tau_v$ the vehicle is at $v$ with object $j$ in it. Consider the tour
$\sigma_v$ obtained by modifying $\tau_v$ so that it drops each object $j$ at $v$ when the vehicle is at $v$ with
object $j$ in it. Clearly $\kappa(\sigma_v)=\kappa(\tau_v)=\lb(v)$.

Note that $\sigma_v$ is a feasible preemptive tour for the {\em single source Dial-a-Ride}~problem with sink $v$ and
all sources in $L_v$. Setting delay parameter $\beta=2$ in Theorem~\ref{th:bnd-delay} gives a {\em non-preemptive} tour
$\sigma'_v$ that moves all objects in $L_v$ from their sources to $v$, having length at most
$5.5\kappa(\sigma_v)=5.5\lb(v)$ such that the time spent by each object in the vehicle is at most twice the distance
from its source to $v$ (which is its nca). Similarly, we can obtain a non-preemptive tour $\sigma''_v$ that moves all
objects in $L_v$ from $v$ to their destinations, having length at most $5.5\lb(v)$, where the time spent by each object
in the vehicle is at most twice the distance from its destination to $v$. Now $\sigma'_v\cdot \sigma''_v$ is a {\em
1-preemptive} tour servicing $L_v$, of length at most $11\cdot \lb(v)$, where each object $i\in L_v$ spends at most
$2(\kappa(s_i,v)+\kappa(t_i,v))=2\cdot\kappa(s_i,t_i)$ time in the vehicle.}

%In Appendix~\ref{app:eucl-mdr}, we show that these bounds can be improved by a logarithmic factor when the underlying metric is a fixed dimensional Euclidean space.

\subsection{Algorithm for Capacitated \mdr\label{subsec:mdar-algo}}
We are now ready to present our algorithm for capacitated multi-vehicle \dr. The algorithm for \mdr relies on a partial
coverage algorithm \cov that given subsets $Q\sse [q]$ of vehicles and $D\sse[m]$ of demands, outputs a schedule for
$Q$ of near-optimal makespan that covers some {\em fraction} of demands in $D$. The main steps in \cov are as follows.
\begin{enumerate}
 \item Obtain a {\em single-vehicle} tour satisfying 1-preemptive and bounded-delay properties
(Theorem~\ref{th:1pmt-bnd-delay}).
 \item Randomly partition the single vehicle tour into $|Q|$ equally spaced pieces.
 \item Solve a matching problem to assign {\em some} of these pieces to vehicles of $Q$, so as to satisfy a subset of
 demands in $D$.
  \item A suitable fraction of unsatisfied demands in $D$ are covered recursively by unused vehicles of $Q$.
\end{enumerate}

The algorithm first guesses the optimal makespan $B$ of the given instance of \mdr (it suffices to know $B$ within a
constant factor, which is required for a polynomial-time algorithm).  Let parameter $\alpha:=1-\frac{1}{1+\lg m}$. For
any subset $P\sse [q]$, we abuse notation and use $P$ to denote both the set of vehicles $P$ and the multi-set of
depots corresponding to vehicles $P$.

We give an algorithm \cov that takes as input a tuple $\langle Q, D, B\rangle$ where $Q\sse [q]$ is a subset of
vehicles, $D\sse [m]$ a subset of demands and $B\in \mathbb{R}_+$, with the {\em promise} that vehicles $Q$
(originating at their respective depots) suffice to completely serve the demands $D$ at a makespan of $B$. Given such a
promise, \cov$\langle Q, D, B\rangle$ returns a schedule of makespan $O(\log n \log m)\cdot B$ that serves a good
fraction of $D$. Algorithm $\cov\langle Q,D,B\rangle$ is given in Figure~\ref{fig:algo-cap}. We set parameter
$\rho=\Theta(\log n\log m)$, the precise constant in the $\Theta$-notation comes from the analysis.

\begin{figure}[!h]
\begin{center}
\framebox[6.3in]{\parbox{6.1in}{
%\begin{small}
{\bf Input:} Vehicles $Q\sse [q]$, demands $D\sse [m]$, bound $B\ge 0$ {\em such that} $Q$ can serve all demands in $D$
at makespan $B$.
\begin{enumerate}
\item[] {\bf Preprocessing}
\item \label{step:algo-mst}If the minimum spanning tree (MST) on vertices $Q$ contains an edge of length greater than $3B$, there is a non-trivial partition $\{Q_1,Q_2\}$ of $Q$ with $d(Q_1,Q_2)>3B$.
For $j\in\{1,2\}$, let $V_j=\{v\in V\mid d(Q_j,v)\le B\}$ and $D_j$ be all demands of $D$ induced on $V_j$. Run in
parallel the schedules from  $\cov\langle Q_1,D_1,B\rangle$ and  $\cov\langle Q_2,D_2,B\rangle$. Assume this is not the
case in the following.
\item[] {\bf Random partitioning}
\item\label{step:algo-1pmt} Obtain single-vehicle 1-preemptive tour $\tau$ using capacity $k$ and serving demands $D$, by applying Theorem~\ref{th:1pmt-bnd-delay}.
\item\label{step:algo-rand-cut} Choose a uniformly random offset $\eta\in [0, \rho B]$ and cut edges of tour $\tau$ at distances $\{p\rho B+\eta \mid p=1,2,\cdots\}$ along the tour to obtain a
set $\mathcal{P}$ of pieces of $\tau$.
\item\label{step:algo-cut-dem} $C''$ is the set of objects $i\in D$ such that $i$ is carried by the vehicle in $\tau$ over some edge that is cut in Step~\eqref{step:algo-rand-cut}; and $C':= D\setminus C''$. Ignore $C''$ objects in the rest of the
algorithm.
\item[] {\bf Load rebalancing}
\item\label{step:algo-match} Construct bipartite graph $H$ with vertex sets $\mathcal{P}$ and $Q$ and an edge between piece $P\in \mathcal{P}$ and depot $f\in Q$ iff $d(f,P)\le 2B$. For any subset $A\sse \mathcal{P}$,
$\Gamma(A)\sse Q$ denotes the neighborhood of $A$ in graph $H$. Let $\mathcal{S}\sse \mathcal{P}$ be any {\em maximal}
set that satisfies $|\Gamma(\mathcal{S})|\le \frac{|\mathcal{S}|}{2}$.
\item \label{step:algo-2mat} Compute a {\em 2-matching} $\pi:\mathcal{P}\setminus \mathcal{S}\rightarrow Q\setminus \Gamma(\mathcal{S})$, i.e. function s.t. $(P,\pi(P))$ is an edge in $H$ for all
$P\in \mathcal{P}\setminus \mathcal{S}$, and the number of pieces mapping to any $f\in Q\setminus \Gamma(\mathcal{S})$
is $|\pi^{-1}(f)|\le 2$.
\item[] {\bf Recursion}
\item\label{step:algo-partn} Define $C_1:= \{i\in C'\mid \mbox{ either }s_i\in \mathcal{S} \textrm{ or } t_i \in \mathcal{S}\}$; and $C_2:= C'\setminus C_1$.
\item\label{step:algo-sched} Run in parallel the {\em recursive} schedule $\cov\langle \Gamma(\mathcal{S}),C_1,B\rangle$ for $C_1$ and the following for $C_2$:
 \begin{enumerate}
 \item Each vehicle $f\in Q\setminus \Gamma(\mathcal{S})$ traverses the pieces $\pi^{-1}(f)$, moving all $C_2$-objects in them from their source to preemption-vertex, and returns to its depot.
 \item Each vehicle $f\in Q\setminus \Gamma(\mathcal{S})$
again traverses the pieces $\pi^{-1}(f)$, this time moving all $C_2$-objects in them from their preemption-vertex to
destination, and returns to its depot.
 \end{enumerate}
\end{enumerate}
{\bf Output:} A schedule of $Q$ of makespan $(16+16\rho)\cdot B$ that serves an $\alpha^{\lg \min\{|Q|,2m\}}$ fraction
of $D$.
%\end{small}
}}
\end{center}
\caption{Algorithm $\cov\langle Q,D,B\rangle$ for capacitated \mdr.\label{fig:algo-cap}}
\end{figure}

\begin{lem}\label{lem:partial}
If there exists a schedule of vehicles $Q$ covering all demands $D$, having makespan at most $B$, then \cov invoked on
$\langle Q, D, B\rangle$ returns a schedule of vehicles $Q$ of makespan at most $(16+16\rho)\cdot B$ that covers at
least an $\alpha^{\lg z}$ fraction of $D$, where $z:= \min\{|Q|,2m\}\le 2m$.
\end{lem}
The final algorithm invokes \cov iteratively until all demands are covered: each time with the entire set $[q]$ of
vehicles, all uncovered demands, and bound $B$. If $D\sse[m]$ is the set of uncovered demands at any iteration,
Lemma~\ref{lem:partial} implies that $\cov\langle[q],D,B\rangle$ returns a schedule of makespan $O(\log m\log n)\cdot
B$ that serves at least $\frac{1}{4}|D|$ demands. Hence a standard set-cover analysis implies that all demands will be
covered in $O(\log m)$ rounds, resulting in a makespan of $O(\log^2 m\log n)\cdot B$.

\medskip

\noindent {\bf Proof of Lemma~\ref{lem:partial}.} We proceed by induction on the number $|Q|$ of vehicles. The base
case $|Q|=1$ is easy: the tour $\tau$ in Step~\eqref{step:algo-1pmt} has length $O(\log^2n)\cdot B\le \rho B$, and
satisfies all demands (i.e. fraction $1$). In the following, we prove the inductive step, when $|Q|\ge 2$.

\paragraph{Preprocessing.} Suppose Step~\eqref{step:algo-mst} applies. Note that $d(V_1,V_2)>B$ and hence there is no demand
with source in one of $\{V_1,V_2\}$ and destination in the other. So demands $D_1$ and $D_2$ partition $D$. Furthermore
in the optimal schedule, vehicles $Q_j$ (any $j=1,2$) only visit vertices in $V_j$ (otherwise the makespan would be
greater than $B$). Thus the two recursive calls to \cov satisfy the assumption: there is some schedule of vehicles
$Q_j$ serving $D_j$ having makespan $B$. Inductively, the schedule returned by \cov for each $j=1,2$ has makespan at
most $(16+16\rho)\cdot B$ and covers at least $\alpha^{\lg c}\cdot |D_j|$ demands from $D_j$, where $c\le
\min\{|Q|-1,2m\}\le z$. The schedules returned by the two recursive calls to \cov can clearly be run in parallel and
this covers at least $\alpha^{\lg z}(|D_1|+|D_2|)$ demands, i.e. an $\alpha^{\lg z}$ fraction of $D$. So we have the
desired performance in this case.

%From the preceding argument, the MST on the
%end-points of $D$ is at most $4|Q|\cdot B$. %Thus the {\em preemptive lower
%bound} for the single vehicle preemptive Dial-a-Ride with demands
%$D$ is at most $5zB$.
%Ravi added the following lines of outline...
\ignore{ We outline briefly the proof strategy: First we show (Lemma~\ref{lem:1-pmt}) how to obtain a single vehicle
1-preemptive tour serving $D$  with small length with an additional property that the total time spent by the demands
in the vehicle is bounded more tightly. We use the latter property to argue (Claim~\ref{cl:cut}) that if this tour is
randomly cut into $q = |Q|$ roughly equal length segments (as in the single depot case), there is only a small (roughly
$\frac{1}{\log m}$-fraction) of demands that are cut across segments. We then show how to route the remaining uncut
demands (an $\alpha$-fraction) in two steps: First we assign the pieces to vehicles using a matching and route the
demands within these pieces as in the previous section in two rounds; the unmatched pieces and vehicles are identified
and we apply the method recursively to finish routing a good fraction of this second set of demands. This will complete
the proof of Lemma~\ref{lem:partial} for this case; details follow.}

\paragraph{Random partitioning.} The harder part of the algorithm is when Step~\eqref{step:algo-mst} does not apply: so the
MST length on $Q$ is at most $3|Q|\cdot B$. Note that when the depots $Q$ are contracted to a single vertex, the MST on
the end-points of $D$ plus the contracted depot-vertex has length at most $|Q|\cdot B$ (the optimal makespan schedule
induces such a tree). Thus the MST on the depots $Q$ along with end-points of $D$ has length at most $4|Q|\cdot B$.
Based on the assumption in Lemma~\ref{lem:partial} and the flow lower bound for \mdr, we have $\sum_{i\in D}
 d(s_i,t_i)\le k|Q|\cdot B$. It follows that for the {\em single vehicle} \dr instance solved in Step~\eqref{step:algo-1pmt}, the Steiner and flow lower-bounds (denoted
$\lb$ in Theorem~\ref{th:1pmt-bnd-delay}) are $O(1)\cdot |Q|B$. Theorem~\ref{th:1pmt-bnd-delay} now implies that (with
high probability) $\tau$ is a 1-preemptive tour servicing $D$, of length at most $O(\log^2 n)|Q|\cdot B$ such that
$\sum_{i\in D} T_i\le O(\log n)\cdot |D| B$, where $T_i$ denotes the total time spent in the vehicle by demand $i\in
D$. The bound on the delay uses the fact that $\max_{i=1}^m d(s_i,t_i) \le B$.

\begin{figure}[h]
\begin{center}
\includegraphics[scale=0.6]{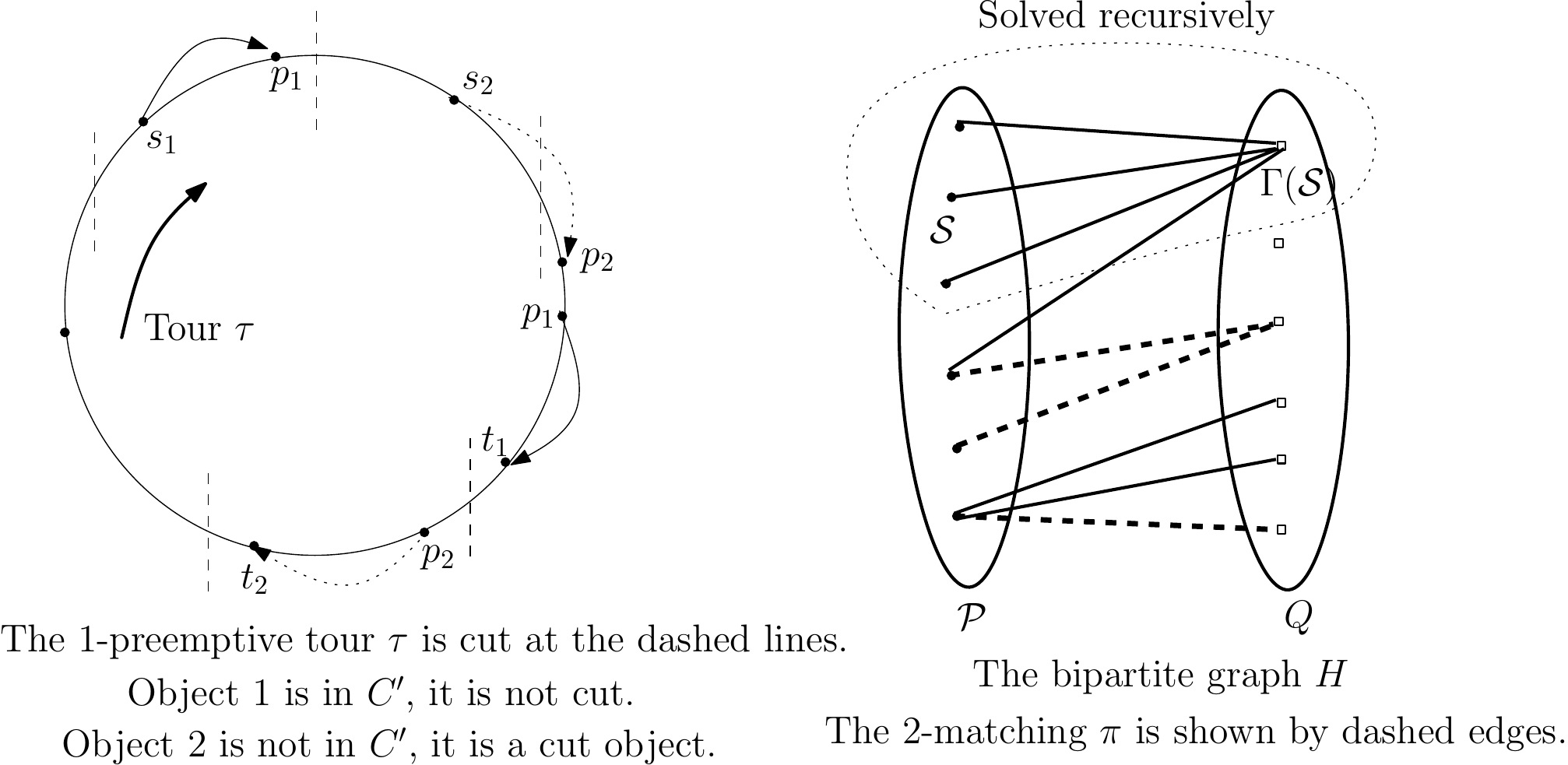}
\end{center}
\caption{\label{fig:cut-patch}Cutting and patching steps in algorithm \cov.}
\end{figure}

Choosing a large enough constant corresponding to $\rho=\Theta(\log n\log m)$, the length of $\tau$ is upper bounded by
$\rho|Q|\cdot B$ (since $n\le 2m$). So the cutting procedure in Step~\eqref{step:algo-rand-cut} results in at most
$|Q|$ pieces of $\tau$, each of length at most $2\rho B$. The objects $i\in C''$ (as defined in
Step~\eqref{step:algo-cut-dem}) are called a {\em cut objects}. We restrict attention to the other objects
$C'=D\setminus C''$ that are not `cut'. For each object $i\in C'$, the path traced by it (under single vehicle tour
$\tau$) from its source $s_i$ to preemption-point and the path from its preemption-point to $t_i$ are both completely
contained in pieces of $\mathcal{P}$. Figure~\ref{fig:cut-patch} gives an example of objects in $C'$ and $C''$, and the
cutting procedure.
\begin{cl}\label{cl:cut}
The expected number of objects in $C''$ is at most $\sum_{i\in D} \frac{T_i}{\rho B}\le O(\frac{1}{\log m})\cdot |D|$.
\end{cl}
\begin{pf}
The probability (over choice of $\eta$) that object $i\in D$ is cut equals $\frac{T_i}{\rho B}$ where $T_i$ is the
total time spent by $i$ in tour $\tau$. The claim follows by linearity of expectation and $\sum_{i\in D} T_i\le O(\log
n)\cdot |D| B$.
\end{pf}

We can derandomize Step~\eqref{step:algo-rand-cut} and pick the best offset $\eta$ (there are at most polynomially many
combinatorially distinct offsets). Claim~\ref{cl:cut} implies (again choosing large enough constant in
$\rho=\Theta(\log n \log m)$) that $|C'|\ge (1-\frac{1}{2\lg m})|D|\ge \alpha\cdot |D|$ demands are {\em not cut}; note
that this occurs with probability one since we use the best choice for $\eta$. Now onwards we only consider the set
$C'$ of uncut demands. Let $\mathcal{P}$ denote the pieces obtained by cutting $\tau$ as above, recall
$|\mathcal{P}|\le |Q|$. A piece $P\in \mathcal{P}$ is said to be non-trivial if the vehicle in the 1-preemptive tour
$\tau$ carries some $C'$-object while traversing $P$. Note that the number of non-trivial pieces in $\mathcal{P}$ is at
most $2|C'|\le 2m$: each $C'$-object appears in at most 2 pieces, one where it is moved from source to
preemption-vertex and other from preemption-vertex to destination. Retain only the non-trivial pieces in $\mathcal{P}$;
so $|\mathcal{P}|\le \min\{|Q|,2m\}=z$. The pieces in $\mathcal{P}$ may not be one-to-one assignable to the depots
since the algorithm has not taken the depot locations into account. We determine which pieces may be assigned to depots
by considering a matching problem between $\mathcal{P}$ and the depots in Step~\eqref{step:algo-match} and
\eqref{step:algo-2mat}.

\paragraph{Load rebalancing.} The bipartite graph $H$ (defined in Step~\eqref{step:algo-match}) represents which pieces and
depots may be assigned to each other. Piece $P\in \mathcal{P}$ and depot $f\in Q$ are assignable iff $d(f,P)\le 2B$,
and in this case graph $H$ contains an edge $(P,f)$. We claim that corresponding to the `maximal contracting' set
$\mathcal{S}$ (defined in Step~\eqref{step:algo-match}), the 2-matching $\pi$ (in Step~\eqref{step:algo-2mat}) is
guaranteed to exist. Note that $|\Gamma(\mathcal{S})|\le \frac{|\mathcal{S}|}{2}$, but $|\Gamma(\mathcal{T})|>
\frac{|\mathcal{T}|}{2}$ for all $\mathcal{T}\supset \mathcal{S}$. For any $T'\sse \mathcal{P}\setminus \mathcal{S}$,
let $\widetilde{\Gamma}(T')$ denote the neighborhood of $T'$ in $Q\setminus \Gamma(\mathcal{S})$. The maximality of
$\mathcal{S}$ implies: for any non-empty $T'\sse \mathcal{P}\setminus \mathcal{S}$,
$\frac{|\mathcal{S}|}{2}+\frac{|T'|}{2}=\frac{|\mathcal{S}\cup T'|}{2}< |\Gamma(\mathcal{S}\cup T')|=
|\Gamma(\mathcal{S})|+|\widetilde{\Gamma}(T')|$, i.e. $|\widetilde{\Gamma}(T')|\ge \frac{|T'|}{2}$. Hence by {\em
Hall's condition}, there is a 2-matching $\pi:\mathcal{P}\setminus \mathcal{S}\rightarrow Q\setminus
\Gamma(\mathcal{S})$. The set $\mathcal{S}$ and 2-matching $\pi$ can be easily computed in polynomial time.

\paragraph{Recursion.} In Step~\eqref{step:algo-partn}, demands $C'$ are further partitioned into two sets: $C_1$ consists of
objects that are {\em either} picked-up {\em or} dropped-off in some piece of $\mathcal{S}$; and $C_2$-objects are
picked-up {\em and} dropped-off in pieces of $\mathcal{P}\setminus \mathcal{S}$. The vehicles $\Gamma(\mathcal{S})$
suffice to serve all $C_1$ objects, as shown below.
\begin{cl}\label{cl:recurse}
There exists a schedule of vehicles $\Gamma(\mathcal{S})$ serving demands $C_1$, having makespan $B$.
\end{cl}
\begin{pf}
Consider the schedule of makespan $B$ that serves all demands $C'=C_1\cup C_2$ using vehicles $Q$: this is implied by
the promise on instance $\langle Q,D,B\rangle$. We claim that in this schedule, no vehicle from $Q\setminus
\Gamma(\mathcal{S})$ moves any $C_1$ object. Suppose (for a contradiction) that the vehicle from depot $f\in Q\setminus
\Gamma(\mathcal{S})$ moves object $i\in C_1$ at some point in this schedule; then it must be that $d(f,s_i)\textrm{ and
} d(f,t_i)\le 2B$. But since $i\in C_1$, at least one of $s_i$ or $t_i$ is in a piece of $\mathcal{S}$, and this
implies that there is some piece $P\in \mathcal{S}$ with $d(f,P)\le 2B$, i.e. $f\in \Gamma(\mathcal{S})$, which is a
contradiction! Thus the only vehicles participating in the movement of $C_1$ objects are $\Gamma(\mathcal{S})$, which
implies the claim.
\end{pf}

In the final schedule, a {\em large fraction} of $C_1$ demands are served by vehicles $\Gamma(\mathcal{S})$, and {\em
all} the $C_2$ demands are served by vehicles $Q\setminus \Gamma(\mathcal{S})$. Figure~\ref{fig:cut-patch} shows an
example of this partition.

{\bf Serving $C_1$ demands.} Based on Claim~\ref{cl:recurse}, the recursive call \cov$\langle
\Gamma(\mathcal{S}),C_1,B\rangle$ (made in Step~\eqref{step:algo-sched}) satisfies the assumption required in
Lemma~\ref{lem:partial}. Since $|\Gamma(\mathcal{S})|\le \frac{|\mathcal{P}|}{2}\le \frac{|Q|}{2}<|Q|$, we obtain
inductively that \cov$\langle \Gamma(\mathcal{S}),C_1,B\rangle$ returns a schedule of makespan $(16+16\rho)\cdot B$
covering at least $\alpha^{\lg y}\cdot |C_1|$ demands of $C_1$, where $y=\min\{|\Gamma(\mathcal{S})|,2m\}$. Note that
$y\le |\Gamma(\mathcal{S})|\le |\mathcal{P}|/2\le z/2$ (recall $|\mathcal{P}|\le z$), which implies that at least
$\alpha^{\lg z-1}|C_1|$ demands are covered.

{\bf Serving $C_2$ demands.} These are served by vehicles $Q\setminus \Gamma(\mathcal{S})$ using the 2-matching $\pi$,
in two rounds as specified in Step~\eqref{step:algo-sched}. This suffices to serve all objects in $C_2$ since for any
$i\in C_2$, the paths traversed by object $i$ under $\tau$, namely $s_i\leadsto p_i$ (its preemption-point) and
$p_i\leadsto t_i$ are contained in pieces of $\mathcal{P}\setminus \mathcal{S}$. Furthermore, since $|\pi^{-1}(f)|\le
2$ for all $f\in Q\setminus \Gamma(\mathcal{S})$, the distance traveled by vehicle $f$ in one round is at most $2\cdot
2(2B+2\rho B)$. So the time taken by this schedule is at most $2\cdot 4(2B+2\rho B)=(16+16\rho)\cdot B$.

The schedule of vehicles $\Gamma(\mathcal{S})$ (serving $C_1$) and vehicles $Q\setminus \Gamma(\mathcal{S})$ (serving
$C_2$) can clearly be run in parallel. This takes time $(16+16\rho)\cdot B$ and covers in total at least
$|C_2|+\alpha^{\lg z - 1}|C_1|\ge \alpha^{\lg z - 1}|C'|\ge \alpha^{\lg z} |D|$ demands of $D$. This completes the
proof of the inductive step of Lemma~\ref{lem:partial}.

\medskip

Using Lemma~\ref{lem:partial} repeatedly as mentioned earlier, we obtain an $O(\log^2m\cdot \log n)$-approximation
algorithm. Using some preprocessing steps (described in the next subsection), we have Theorem~\ref{th:p-mdr}.

%---------------------Weighted preemptive DaR-----------------------------

\subsection{Weighted \mdr}\label{subsec:wt-mdar} The multi-vehicle \dr problem as
defined  assumes that all objects have the same `weight', i.e. each object occupies a unit capacity. In the weighted
\mdr problem, each object $i\in [m]$ also has a weight $w_i$, and the capacity constraint requires that no vehicle
carry a total weight of more than $k$ at any time. In this section, we obtain an $O(\log^3n)$-approximation algorithm
for weighted \mdr. In particular this would imply Theorem~\ref{th:p-mdr}.

The algorithm for weighted \mdr first guesses the optimal makespan $B$ of the given instance. It serves the demands in
two phases: the 1st phase involves pre-processing `heavy demands' and has a makespan of $O(1)\cdot B$; the 2nd phase is
identical to the algorithm \cov in Section~\ref{subsec:mdar-algo} and covers all remaining demands. For every $u,v\in
V$ let $\dem_{u,v}$ denote the total weight of objects having source $u$ and destination $v$. Define $H=\{(u,v)\in
V\times V\mid \dem_{u,v}\ge k/2\}$ to be the {\em heavy} vertex-pairs, and $\hat{H}$ the set of demands between pairs
of $H$.

\paragraph{Phase I.} In this pre-processing step, we cover all demands $\hat{H}$ between heavy vertex-pairs.
The algorithm \pre for solving instance $\mathcal{U}$ and its analysis follow along the lines of algorithm \cov in
Section~\ref{subsec:mdar-algo}; in fact it is much easier. Below we use the same notation $P\sse [q]$ for a set $P$ of
vehicles and the multi-set of depots corresponding to $P$.
%For a subset $D\sse H$ of heavy vertex-pairs, let $\hat{D}$ denote the set of demands between pairs of $D$. We describe an algorithm \pre$\langle Q,D\rangle$ that given any subset
%$Q$ of vehicles and $D$ of heavy vertex-pairs, serves all the demands $\hat{H}$.
The algorithm \prew~ will satisfy the following property.
\begin{lem}\label{lem:wt-mdr}
Given any subset $Q\sse [q]$ of vehicles and $D\sse H$ of heavy vertex-pairs and bound $B$ {\em such that} there is a
schedule of vehicles $Q$ covering demands $\hat{D}$ with makespan at most $B$, \prew$\langle Q,D,B\rangle$ returns a
schedule of vehicles $Q$ covering demands $\hat{D}$ that has makespan $O(1)\cdot B$.
\end{lem}
{\bf Note:} Invoking $\pre\langle [q], H, B\rangle$ gives the desired schedule covering $\hat{H}$ at makespan
$O(1)\cdot B$.
%; we provide an outline and highlight the differences.

\smallskip

\ignore{Lemma~\ref{lem:wt-mdr} is proved by induction on $|Q|$. The base case $|Q|=1$ corresponds to single vehicle \dr
with unit capacity. This is the {\em Stacker-crane} problem~\cite{fhk}, for which there is a $1.8$-approximation
relative to the flow and Steiner lower-bounds. Observe that the flow lower-bound equals $\sum_{(u,v)\in H} g_{u,v}\cdot
d(u,v)\le \frac13 \sum_{(u,v)\in H} \,\lceil \frac{\dem_{u,v}}k\rceil \, d(u,v)$, by definition of heavy vertex-pairs.
Since there is a tour of length $B$ (by assumption in Lemma~\ref{lem:wt-mdr}), it must be that $\sum_{(u,v)\in H}
\,\lceil \frac{\dem_{u,v}}k\rceil \, d(u,v)\le 3B$. Consider the following solution serving demands $\hat{D}$: the
vehicle follows a $1.5$-approximate TSP tour on the depot and all the sources in $D$, and when visiting the source $u$
of any vertex-pair $(u,v)\in D$, the vehicle performs

In the following we consider the inductive step (where $|Q|\ge 2$).}

Algorithm \pre first ensures that each edge of the minimum spanning tree on $Q$ has length at most $3B$, otherwise the
problem decouples into two disjoint smaller problems (as in Step~\eqref{step:algo-mst} of \cov). Hence, as in algorithm
\cov, we also obtain that:
\begin{equation} \label{eq:wt-pre-st} \mbox{ The MST on all sources/destinations in $D$ has length at most
} O(1)\cdot |Q|B\end{equation}

Demands with source $u$ and destination $v$ are referred to as $(u,v)$ demands. For every $(u,v)\in D$, using a greedy
procedure, one can partition all $(u,v)$ demands such that the total weight of each part (except possibly the last) is
between $\frac{k}{2}$ and $k$. For any $(u,v)\in D$, let $g_{u,v}$ denote the number of parts in this partition of
$(u,v)$ demands; note that $g_{u,v}\le \frac{2}{k}\dem_{u,v}+1 \le \frac{4}{k}\dem_{u,v}$ (since $\dem_{u,v}\ge k/2$).
The flow lower bound for the \mdr instance with vehicles $Q$ and demands $\hat{D}$ equals $\frac{1}{|Q|k}\sum_{(u,v)\in
D} \dem_{u,v}\cdot d(u,v)\ge \frac{1}{4|Q|} \sum_{(u,v)\in D} g_{u,v}\cdot d(u,v)$. Using the assumption in
Lemma~\ref{lem:wt-mdr}, we have that:
\begin{equation}\label{eq:wt-pre-flow}
\sum_{(u,v)\in D} g_{u,v}\cdot d(u,v)\le 4|Q|\,B. \end{equation}

In the rest of algorithm \prew, we will consider each part in the above partition of $(u,v)$ demands (for each
$(u,v)\in D$) as a single object of weight $k$; so all the objects in one part will be moved together. Scaling down the
weights and capacity by $k$, we obtain the following equivalent {\em unit-weight unit-capacity instance} $\mathcal{U}$:
for each $(u,v)\in D$ there are $g_{u,v}$ demands with weight 1, source $u$ and destination $v$, and each vehicle in
$Q$ has capacity 1.
%Let $[m']$ denote the index-set of the demands in $\mathcal{U}$.

%The flow lower bound of the demands $\hat{H}$ in the original instance implies that $\frac{1}{4|Q|} \sum_{(u,v)\in H}
%g_{u,v}\cdot d(u,v)\le B$; i.e. the flow bound in $\mathcal{U}$ is at most $4|Q|B$.
Since $\mathcal{U}$ has unit capacity and weights, we can use the {\em Stacker-crane} algorithm~\cite{fhk} to obtain a
single vehicle {\em non-preemptive} tour $\tau$ serving all demands, having length $1.8$ times the Steiner and flow
lower bounds. For this single vehicle problem, the lower bounds corresponding to $\mathcal{U}$ are: (Steiner bound) TSP
on all sources/destinations in $D$, and (flow bound) $\sum_{(u,v)\in H} g_{u,v}\cdot d(u,v)$. From~\eqref{eq:wt-pre-st}
and~\eqref{eq:wt-pre-flow} above, these are both $O(1)\cdot |Q|B$. The algorithm next cuts tour $\tau$ to obtain at
most $|Q|$ pieces $\mathcal{P}$ such that each piece has length $O(1)\cdot B$. In addition it can be ensured that {\em
no} object is being carried over the cut edges: since the vehicle carries at most one object at any time and each
source-destination distance is at most $B$. When $|Q|=1$, the algorithm ends here, and returns this single tour of
length $O(1)\,B$ that serves all demands $\hat{D}$. This proves the base case of Lemma~\ref{lem:wt-mdr}.

When $|Q|\ge 2$, construct bipartite graph $\mathcal{H}$ with vertex sets $\mathcal{P}$ and $Q$ and an edge between
piece $P\in \mathcal{P}$ and depot $f\in Q$ iff $d(f,P)\le 2B$; $\Gamma(A)$ denotes the neighborhood of any $A\sse
\mathcal{P}$ in $\mathcal{H}$. The algorithm finds any maximal set $\mathcal{S}\sse \mathcal{P}$  with
$|\Gamma(\mathcal{S})|< |\mathcal{S}|/2$ (as in Step~\eqref{step:algo-match} of \cov). This implies a 2-matching
$\pi:\mathcal{P}\setminus \mathcal{S}\rightarrow Q\setminus \Gamma(\mathcal{S})$ such that there exists a schedule of
vehicles $\Gamma(\mathcal{S})$ serving demands in the pieces $\mathcal{S}$ with makespan $B$ (c.f.
Claim~\ref{cl:recurse}). Let $C\sse D$ denote the heavy demand-pairs served in some piece of $\mathcal{S}$. The final
schedule returned by $\pre\langle Q,D\rangle$ involves: (i) schedule for vehicles $Q\setminus \Gamma(\mathcal{S})$
given by $\pi$ (covering demand-pairs $D\setminus C$); and (ii) schedule for vehicles $\Gamma(\mathcal{S})$ obtained
recursively $\pre\langle \Gamma(\mathcal{S}),C\rangle$ (covering demands $C$). The recursively obtained schedule has
makespan $O(1)\cdot B$ by the induction hypothesis (since $|\Gamma(\mathcal{S})|<|Q|$, and $\langle
\Gamma(\mathcal{S}),C\rangle$ satisfies the assumption in Lemma~\ref{lem:wt-mdr}); hence the final schedule also has
makespan $O(1)\cdot B$, which proves the inductive step.

\paragraph{Phase II.} Let $L=\{(u,v)\in V\times V\mid 0<\dem_{u,v}< k/2\}$ be the {\em light} vertex-pairs.
The algorithm for this phase treats all $(u,v)$ demands as a {\em single object} of weight $\dem_{u,v}$ from $u$ to
$v$; so there are $m=|L|\le n^2$ distinct objects. The algorithm for this case is identical to \cov of
Section~\ref{subsec:mdar-algo} for the unweighted case: Theorem~\ref{th:1pmt-bnd-delay} generalizes easily to the
weighted case, and the Steiner and flow lower-bounds stay the same after combining demands in $L$. Thus we obtain an
$O(\log^2m\,\log n)=O(\log^3n)$ approximate schedule that covers all remaining demands (setting $m\le n^2$).

\begin{thm}
There is a randomized $O(\log^3n)$-approximation algorithm for weighted preemptive \mdr.
\end{thm}

%-----------------------Excluded minor---------------------

\subsection{Improved Guarantee for Metrics Excluding a Fixed Minor}\label{subsec:cap-planar}

In this section, we show that our algorithm for preemptive \mdr achieves an $O(\log^2 n)$ approximation guarantee when
the metric is induced by a graph excluding any fixed minor.  The main ingredient is the following improvement in the
single vehicle structure from Theorem~\ref{th:1pmt-bnd-delay}.

\begin{thm}\label{th:1pmt-bnd-delay-planar}
There is a randomized poly-time  algorithm that for any instance of single vehicle dial-a-ride on metrics excluding a
fixed minor, computes a 1-preemptive tour $\tau$ servicing all demands $D$ that satifies the following conditions
(where $\lb$ is the average of the Steiner and flow lower bounds):
\begin{enumerate}
  \item {\bf Total length:} $d(\tau)\le O(\log n)\cdot \lb$.
  \item {\bf Bounded delay:} $\sum_{i\in D} T_i\le O(1)\sum_{i\in D} d(s_i,t_i)$ where
$T_i$ is the total time spent by object $i\in D$ in the vehicle under the schedule given by $\tau$.
\end{enumerate}
\end{thm}
Using this 1-preemptive tour within the algorithm of Section~\ref{subsec:mdar-algo} (setting parameter
$\rho=\Theta(\log m)$), we immediately obtain Theorem~\ref{th:cap-planar}.
%and Theorem~\ref{th:cap-eucl}.
In proving Theorem~\ref{th:1pmt-bnd-delay-planar}, we first show that metrics induced by (fixed) minor-free graphs
allow a so-called \emph{$\gamma$-separated cover} (Subsection~\ref{subsub:sep-cov}). Then we show (in
Subsection~\ref{subsub:1pmt-planar}) how this property can be used to obtain the single-vehicle tour  claimed in
Theorem~\ref{th:1pmt-bnd-delay-planar}.

\subsubsection{$\gamma$-Separated Covers}\label{subsub:sep-cov}
\def\coll{\ensuremath{\mathcal{C}}\xspace}

We are given a metric $(V,d)$ that is induced by a graph $G=(V,E)$ and edge-lengths $w:E\rightarrow \mathbb{Z}_+$. For
any pair $u,v\in V$ of vertices, the distance $d(u,v)$ equals the shortest path between $u$ and $v$ under edge-lengths
$w$. We assume that the graph $G$ does not contain any $K_r$-minor; here $r$ is a fixed parameter.
%We will show that
Recall that a {\em cluster} refers to any subset of vertices. We prove the following.

\begin{thm}\label{th:sep-cover}
Given a $K_r$-minor free graph $G=(V,E,w)$ and integer $\gamma\ge 0$, there is a polynomial-time algorithm to compute a
collection \coll of clusters along with a partition $Z_1, \ldots, Z_p$ of \coll such that:
%is called a iff the following hold:
\begin{enumerate}
 \item \label{step:sep-cov1} $p= 2^{O(r)}$.
 \item \label{step:sep-cov2} For each $l\in[p]$ and distinct clusters $A, B \in Z_l$, the distance between $A$ and $B$, $d(A,B)\ge \gamma$.
 \item \label{step:sep-cov3}The diameter of any cluster in $S\in\coll$, $\max_{u,v\in S} d(u,v) \le O(r^2)\cdot \gamma$.
 \item \label{step:sep-cov4}For any pair $u, v\in V$ of vertices with $d(u,v)\leq \gamma$, there is some cluster $A \in \coll$ having $u,v \in A$.
\end{enumerate}
Such a collection \coll is called $\gamma$-separated cover.
\end{thm}

The proof of this theorem essentially follows from the KPR decomposition algorithm~\cite{kpr}. In fact without
conditions (1-2), it is implied by Theorem~\ref{th:sparse-cover}. Achieving conditions (1-2) requires a further
modification to the KPR decomposition, as described below.

%In Section~\ref{sec:planarmdr} and Section~\ref{sec:euclidmdr} we show that planar graphs and constant dimension
%Euclidean metrics, respectively, allows a $\gamma$-separated sparse cover.

We first remind the reader of the main property of the KPR decomposition~\cite{kpr}. We will consider graph $G$ as
having unit length edges, by subdividing each edge $(u,v)\in E$ to become a path of length $w_{uv}$ between $u$ and $v$
(this also increases the number of vertices). The distance function $d$ remains unchanged by this modification.
%In the rest of this discussion,
\begin{defn}[Theorem~4.2, \cite{kpr}]\label{def:kpr}Let $G_1=G$ and $\delta\in \mathbb{Z}_+$ a
distance parameter. Let $G_1,G_2,\cdots,G_{r+1}$ be any sequence of subgraphs of $G$, where each $G_{i+1}$ is obtained
from $G_i$ as follows:
\begin{enumerate}
\item Construct a breadth-first-search tree in $G_i$ from an arbitrary root-vertex.
\item Select any  set of $\delta$ consecutive levels in this breadth-first-search tree, and let $G_{i+1}$ be any
connected component of the subgraph in $G$ induced by these levels.
\end{enumerate}
Then, the diameter of $G_{r+1}$ is $O(r^2)\cdot \delta$.
\end{defn}

\def\spl{\ensuremath{{\sf Split}}\xspace}

We are now ready to proceed with the proof of Theorem~\ref{th:sep-cover}. We give a recursive procedure to
%algorithm \spl that
generate the desired $\gamma$-separated cover \coll of $G$. We initialize $\coll=\emptyset$, and invoke $\spl\langle
G,V,0,\phi\rangle$. Algorithm \spl takes as input: an induced subgraph $H=(V(H),E(H))$ of $G$, {\em terminals} $T\sse
V(H)$, depth of recursion $i$, and {\em color} $\tau\in \{0,1,2\}^{i}$. Then it does the following.
\begin{enumerate}
 \item If $i=r$ then add tuple $\langle T,\tau\rangle$ to collection \coll, and stop.
 \item Construct a BFS tree from any vertex of $H$, and enumerate the levels starting from zero (at the root).
 \item \label{step:well-sep} For each $j\in\{0,1,2\}$, integer $l\ge 0$, and connected component $H'$ in the subgraph of $G$
 induced by levels $\{(3 l+j-1)\gamma,\cdots, (3 l+j+3)\gamma\}$ of BFS do:
 \begin{enumerate}
  \item Set $T'\leftarrow$ vertices of $T\cap V(H')$ in levels $\{(3 l+j)\gamma,\cdots, (3 l+j+2)\gamma\}$ of BFS.
  \item Set $\tau'\leftarrow$ $\tau$ concatenated with $j$.
  \item Recurse on $\spl\langle H',T',i+1,\tau'\rangle$.
 \end{enumerate}
\end{enumerate}
At the end of the algorithm, \coll consists of several tuples of the form $\langle C,\tau\rangle$ where $C$ is a
cluster and $\tau\in \{0,1,2\}^r$ is its color. This naturally corresponds to partitioning \coll into $p=3^r$ parts:
clusters of the same color form a part. We now show that \coll satisfies the conditions in Theorem~\ref{th:sep-cover}.
Condition~1 clearly holds by the construction.

{\bf Condition~4.} Let $u,v\in V$ be such that $d(u,v)\le \gamma$, and let $P\sse V$ denote the vertices on any
shortest $u-v$ path in $G$ (note that $|P|\le \gamma+1$). We claim inductively that for each $i\in \{0,\cdots,r\}$,
there is some call to \spl at depth $i$ such that all vertices of $P$ appear as terminals. The base case $i=0$ is
obvious since all vertices are terminals. Assuming the claim to be true for depth $i$, we prove it for $i+1$. Let
$\langle H,T,i,\tau\rangle$ denote the call to \spl at depth $i$ where $P\sse T\sse V(H)$. Since the $P$-vertices
define a path of length at most $\gamma$ (in $G$ and also $H$), all $P$-vertices appear in some $\gamma$ consecutive
levels of any BFS on $H$. Thus there exist values $j\in\{0,1,2\}$ and $l\in \mathbb{Z}_+$ such that $P$ is contained in
levels $\{(3l+j)\gamma,\cdots, (3 l+j+2)\gamma\}$ of the BFS on $H$. This implies that one of the recursive calls
(corresponding to this value of $j$ and $l$) contains $P$ as terminals. Using this claim when $i=r$ implies that there
is some cluster in \coll containing both $u$ and $v$ (in fact containing the entire path $P$).

{\bf Condition 3.} This is a direct consequence of the KPR decomposition. Note that every subgraph at depth $r$ of
algorithm \spl is obtained from $G$ via the sequence of operations in Definition~\ref{def:kpr} with $\delta=4\gamma+1$.
Thus each such subgraph has diameter $O(r^2)\cdot \gamma$. Finally, any cluster in \coll is a subset of some subgraph
at depth $r$ of \spl; so it also has diameter $O(r^2)\cdot \gamma$.

{\bf Condition 2.} This well-separated property is the main reason for the use of `terminals' in algorithm \spl and the
decomposition defined in Step~\eqref{step:well-sep}.  Let $A,B$ be any two distinct clusters in \coll having the {\em
same color} : we will show that $d(A,B)\ge \gamma$. Each of $A$ and $B$ corresponds to a ``root-leaf path'' in the
recursion tree for algorithm \spl (where the root is $\spl\langle G,V,0,\phi\rangle$). Consider the call $\langle
H,T,i,\tau\rangle$ to \spl after which the paths for $A$ and $B$ diverge; let $\alpha_a=\langle
H_a,T_a,i+1,\tau'\rangle$ and $\alpha_b=\langle H_b,T_b,i+1,\tau'\rangle$ denote the respective recursive calls from
$\langle H,T,i,\tau\rangle$ where $A\sse T_a$ and $B\sse T_b$ (note that both have the same color $\tau'$ since $A$ and
$B$ have the same color at the end of the algorithm). We will show that $d(T_a,T_b)\ge \gamma$ which in particular
implies $d(A,B)\ge \gamma$. Note that both $\alpha_a$ and $\alpha_b$ are generated by the same value $j\in\{0,1,2\}$,
since they have the same color. Let $\alpha_a$ (resp. $\alpha_b$) correspond to value $l=l_a$ (resp. $l=l_b$). Consider
the following cases:
\begin{enumerate}
 \item $l_a\ne l_b$. In this case, we show that the terminal-sets $T_a$ and  $T_b$ are far apart in the BFS on $H$.
 Observe that $T_a$ appears within levels $\{(3 l_a+j)\gamma,\cdots, (3 l_a+j+2)\gamma\}$; and $T_b$ within $\{(3 l_b+j)\gamma,\cdots, (3
 l_b+j+2)\gamma\}$. If $l_a<l_b$ (the other case is identical), then $3l_b+j\ge 3l_a+j+3$; i.e. $T_a$ and $T_b$ are
 separated by at least $\gamma-1$ levels in the BFS of $H$. Using Claim~\ref{cl:well-sep} below,
 we have $d(T_a,T_b)\ge \gamma$ in $G$; otherwise $H$ contains a path from $T_a$ to $T_b$ of length at most $\gamma-1$, meaning that
 $T_a$ and $T_b$ are separated by $\le \gamma-2$ levels in the BFS, a contradiction!

 \item $l_a = l_b=l$. In this case, it must be that $H_a$ and $H_b$ are two {\em disconnected components} of the subgraph (of
 $G$) induced on levels $\{(3 l+j-1)\gamma,\cdots, (3 l+j+3)\gamma\}$ of the BFS on $H$. Furthermore, $T_a$ and $T_b$
 both appear within levels $\{(3 l+j)\gamma,\cdots, (3 l+j+2)\gamma\}$. Again by Claim~\ref{cl:well-sep}, we must have $d(T_a,T_b)\ge \gamma$ in
 $G$; otherwise $H$ contains a path from $T_a$ to $T_b$ of length at most $\gamma-1$, which would mean that $H_a$ and
 $H_b$ are connected within levels $\{(3 l+j-1)\gamma,\cdots, (3 l+j+3)\gamma\}$.
\end{enumerate}

\begin{claim}\label{cl:well-sep}
For every call $\spl\langle H,T,i,\tau\rangle$, and terminals $x,y\in T$, if $d(x,y)\le \gamma$ (in graph $G$) then $H$
contains an $x-y$ path of length at most $d(x,y)$.
\end{claim}
\begin{pf}
We proceed by induction on $i$. The claim is obviously true for the call $\spl\langle G,V,0,\phi\rangle$ at depth $0$.
Assuming the claim for any depth $i$ call $\spl\langle H,T,i,\tau\rangle$, we prove it for any call $\spl\langle
H',T',i+1,\tau'\rangle$ generated by it. Let $x,y\in T'$ with $d(x,y)\le \gamma$. Clearly $x,y\in T$, and by the
induction hypothesis, $H$ contains an $x-y$ path $\pi$ of length at most $d(x,y)\le \gamma$. It suffices to show that
$\pi$ is also contained in $H'$. Let $j\in\{0,1,2\}$ and $l\in \mathbb{Z}_+$ denote the values that generated the
recursive call $\spl\langle H',T',i+1,\tau'\rangle$. Since $x,y\in T'$, both these vertices lie within levels $\{(3 l+j
)\gamma,\cdots, (3 l +j+2)\gamma\}$ of the BFS on $H$. Furthermore, $\pi$ is an $x-y$ path in $H$ of length at most
$\gamma$; so $\pi$ is contained within levels $\{(3 l+j-1)\gamma,\cdots, (3 l +j+3)\gamma\}$ and so it lies in the
connected component $H'$ that contains $x$ and $y$.
\end{pf}

\medskip

\noindent {\bf Running time.} It is easy to see that each call to \spl generates only polynomial number of recursive
calls. Since the depth of the recursion is at most $r$, the running time is polynomial for fixed $r$.

\smallskip

\noindent This completes the proof of Theorem~\ref{th:sep-cover}.

\medskip

We note that the collection \coll also satisfies the `sparse cover' property (i.e. condition~(3) in
Theorem~\ref{th:sparse-cover}), namely each vertex $v\in V$ appears in at most $O(2^{r})$ clusters. It is easy to show
(inductively) that the number of depth $i$ calls to \spl where any vertex $v$ appears as a terminal is at most $2^i$.
However this property is not required in the following proof of Theorem~\ref{th:1pmt-bnd-delay-planar}.

\subsubsection{Algorithm for Theorem~\ref{th:1pmt-bnd-delay-planar}}\label{subsub:1pmt-planar}
Recall that we are given an instance of single-vehicle \dr on a metric $(V,d)$ induced by an edge-weighted $K_r$-minor
free graph $G=(V,E,w)$; here $w:E\rightarrow \mathbb{Z}_+$ denotes the edge-lengths and $d$ the shortest-path distances
under $w$. The vehicle has capacity $k$ and $\{s_i,t_i\}_{i\in D}$ are the demand-pairs. Again $r$ is a fixed constant.
\lb denotes the average of the Steiner (i.e. minimum TSP tour on all sources and destinations) and flow lower-bounds
(i.e. $\frac1k \sum_{i\in D} d(s_i,t_i)$) for this instance.

Let $\Delta$ denote the diameter of the metric; by standard scaling arguments, we may assume WLOG that $\Delta$ is at
most polynomial in $n=|V|$. We group the demands $D$ into into $\lceil\log_2 \Delta\rceil$ groups based on the
source-destination distances: For each $j=1,\cdots ,\lceil \log_2 \Delta\rceil$, group $G_j$ consists of those demands
$i\in D$ with $2^{j-1}\le d(s_i,t_i)\le 2^j$. We will show how to service each group $G_j$ separately at a cost of
$O(1)\cdot \lb$ such that the time spent by each object $i\in G_j$ in the vehicle is $T_i=O(2^j) = O(1)\cdot
d(s_i,t_i)$. Since the number of groups is $O(\log n)$, this would imply the theorem.

%In the following, we restrict to some group $G_j$ and let \\

\paragraph{Serving group $G_j$} Set distance parameter $\gamma=2^j$, and obtain a $\gamma$-separated cover \coll along with
its partition $\{Z_1,\cdots,Z_p\}$, using Theorem~\ref{th:sep-cover}. Each demand $i\in G_j$ is {\em assigned} to some
cluster of \coll
%$\cup_{l=1}^p Z_l$
containing both its end-points: since $d(s_i,t_i)\le \gamma$,  this is well-defined by property~\eqref{step:sep-cov4}
of Theorem~\ref{th:1pmt-bnd-delay-planar}. We further partition demands in $G_j$ into $H_1,\cdots,H_p$ where each $H_l$
contains all demands assigned to clusters of $Z_l$. The algorithm will serve demands in each $H_l$ separately by a tour
of length $O(1)\cdot \lb$ that also satisfies the bounded delay condition~2. This suffices to prove the theorem since
$p=O(1)$ by property~\eqref{step:sep-cov1} of Theorem~\ref{th:sep-cover}.

\paragraph{Serving $H_l$.}
For any cluster $S\in Z_l$ let $B(S)\sse G_j$ denote the demands assigned to $S$. Let $\lb(S)$ denote the average of
the Steiner and flow lower bounds restricted to demands $B(S)$. Any cluster $S\in Z_l$ is called non-trivial if
$B(S)\ne \emptyset$. Also pick an arbitrary vertex in each cluster as its {\em center}. Let $\mathcal{T}$ denote a
1.5-approximate TSP tour containing the centers of all non-trivial clusters of $Z_l$; this can be computed in
polynomial time~\cite{c}.
\begin{claim}
\label{cl:small-squares} We have $\sum_{S\in Z_l} \lb(S) \le O(1)\cdot \lb$, and
%the length of tour $\mathcal{T}$ is
$d(\mathcal{T})\le O(1)\cdot \lb$.
\end{claim}
\begin{pf}
It is clear that the sum of the flow lower-bounds over all clusters $S\in Z_l$ is at most the flow lower-bound on
demands $H_l\sse D$. Let $\tsp(S)$ denote the minimum length TSP on the end points of demands $B(S)$, i.e. the Steiner
lower bound on $S$. We show below that $\sum_{S\in Z_l} \tsp(S) \le O(1)\cdot \tsp$ where $\tsp$ denotes the minimum
TSP on end points of $H_l\sse D$ (i.e. Steiner lower bound on $H_l$). This would prove the first part of the claim.

For any cluster $S\in Z_l$,
%consider the TSP tour $\tsp$ restricted to $S$.
let $n(S)$ denote the number of connected segments in $\tsp \cap S$ (i.e. $\tsp$ restricted to $S$), and $C^S_1,\cdots
,C^S_{n(S)}$ denote these segments. Note that tour $\tsp$ travels from one cluster in $Z_l$ to another at least
$\sum_{S\in Z_l} n(S)$ times. By property~\eqref{step:sep-cov2} of Theorem~\ref{th:1pmt-bnd-delay-planar}, the distance
between any two distinct clusters in $Z_l$ is at least $\gamma$. So we obtain:
\begin{equation}\label{eq:tsp-clusters}
\tsp\ge \gamma\sum_{S\in Z_l} n(S)
\end{equation}
We define a TSP tour on end-points of demands $B(S)$ by connecting the end points of segments $C^S_1,\cdots
,C^S_{n(S)}$. This requires $n(S)$ new edges, each of length at most $O(\gamma)$ (i.e. diameter of cluster $S$, by
property~\eqref{step:sep-cov3} of Theorem~\ref{th:1pmt-bnd-delay-planar}). So the minimum TSP tour for any $S\in Z_l$
has length $\tsp(S)\le \sum_{a=1}^{n(S)} d(C^S_a)+ O(\gamma)\cdot n(S)$. Hence,
$$\sum_{S\in Z_l} \tsp(S) \le
\sum_{S\in Z_l}\sum_{a=1}^{n(S)} d(C^S_a)+ O(\gamma) \sum_{S\in Z_l}n(S)\le \tsp +O(\gamma) \sum_{S\in Z_l}n(S)\le
O(1)\cdot\tsp \;.$$

Above, the second inequality follows from the fact that the $C^S_l$s are disjoint segments in $\tsp$, and the last
inequality uses~\eqref{eq:tsp-clusters}.
%Thus we have the first part of the claim.
%This implies that $\sum_{S\in Z_l} \lb(S) \le O(1)\cdot \lb$.

To obtain the second part of the claim, augment $\tsp$ as follows. For each non-trivial cluster $S\in Z_l$,  add two
edges to the center of $S$ from some vertex in $\tsp \cap S$. This gives a TSP tour visiting the centers of all
non-trivial clusters in $Z_l$. Note that the number of non-trivial clusters of $Z_l$ is at most $\sum_{S\in Z_l} n(S)$
since $n(S)\ge 1$ for all non-trivial $S\in Z_l$. Since the diameter of each $S\in Z_l$ is $O(\gamma)$
(property~\eqref{step:sep-cov3} of Theorem~\ref{th:1pmt-bnd-delay-planar}),  the increase in length of $\tsp$ is at
most $2\cdot O(\gamma) \sum_{S\in Z_l} n(S)\le O(1)\cdot \tsp$ by~\eqref{eq:tsp-clusters}. So there is a TSP tour on
the centers of $Z_l$'s non-trivial clusters having length at most $O(1)\cdot \lb$.
\end{pf}

For each $S\in Z_l$, the demands $B(S)$ assigned to it are served separately.
%Let $b(S)$ denote the number of demands assigned to $S$, then
The flow lower-bound on $S$ is at least $\frac{\gamma}{2k} |B(S)|$ (recall that each demand $i\in G_j$ has
$d(s_i,t_i)\ge \gamma/2$).
%Let $C(S)$ denote a 1.5-approximate TSP tour on the end-points of demands in $S$; note that $C(S)$ has length at most $1.5$ times the Steiner lower bound.
Consider an instance $\mathcal{I}_{src}$ of CVRP with all sources from $B(S)$ and $S$'s center as the common
destination. Since the diameter of $S$ is $O(\gamma)$, the flow lower bound of $\mathcal{I}_{src}$ is at most
$O(\gamma) \cdot \frac{|B(S)|}{k}\le O(1)\cdot \lb(S)$. The Steiner lower bound of $\mathcal{I}_{src}$ is also
$O(1)\cdot \lb(S)$. So Theorem~\ref{th:bnd-delay} (with delay parameter $\beta=2$) implies a non-preemptive tour
$\tau_{src}(S)$ that moves all objects from sources of $B(S)$ to the center of $S$, having length $O(1)\cdot \lb(S)$
such that each object spends at most $O(1)\cdot \gamma$ time in the vehicle. Similarly, we can obtain non-preemptive
tour $\tau_{dest}(S)$ that moves all objects from the center of $S$ to destinations of $B(S)$, having length $O(1)\cdot
\lb(S)$ such that each object spends at most $O(1)\cdot \gamma$ time in the vehicle. Concatenating $\tau_{src}(S)$ and
$\tau_{dest}(S)$ gives a 1-preemptive tour serving demands in $S$ having length $O(1)\cdot \lb(S)$ such that each
object $i\in B(S)$ spends at most $O(\gamma) \le O(1)\cdot d(s_i,t_i)$ time in the vehicle.

The final solution for demands $H_l$ traverses the TSP tour $\mathcal{T}$ on centers of all non-trivial clusters of
$Z_l$, and serves the demands in each cluster (as above) when it is visited. The resulting solution has length at most
$d(\mathcal{T}) + O(1) \sum_{S\in Z_l} \lb(S) \le O(1)\cdot \lb$, by Claim~\ref{cl:small-squares}. Moreover,  each
object $i\in H_l$ spends at most $O(1)\cdot d(s_i,t_i)$ time in the vehicle.

\medskip

Finally, concatenating the tours serving each $H_l$ (for $l=1,\cdots,p$), and then the tours serving each $G_j$ (for
$j=1,\cdots,\lceil \log_2\Delta\rceil$),  we obtain the 1-preemptive tour claimed in
Theorem~\ref{th:1pmt-bnd-delay-planar}.

\bibliography{mdar}
\bibliographystyle{abbrv}

\end{document}